\documentclass{revtex4}

\usepackage{graphicx}
\def\underlinewords#1{%
	\def\stuff{#1 }\leavevmode\expandafter\ulword\stuff * }
	\def\ulword#1 {\def\one{#1} \ifx\one\aster\let\next\relax
	\else\vtop{\hbox{\strut#1}\hrule\relax}
	\let\next\ulword\fi\next}
\def\strikeoutwords#1{%
	\def\stuff{#1 }\leavevmode\expandafter\soword\stuff * }
	\def\soword#1 {\def\one{#1} \ifx\one\aster\let\next\relax
	\else\vtop{\hbox{\strut#1}\kern-.5\baselineskip\hrule\relax}
	\let\next\soword\fi\next}
\def\aster{*}

\begin{document}

\begin{flushright}
{$
\begin{array}{l}
\mbox{UB-HET-02-01}\\
%\mbox{hep-ph/9802xxx} \\
\mbox{January~2002} \\ [5mm]
\end{array}
$}
\end{flushright}

% You should use BibTeX and revtex.bst for references
\bibliographystyle{revtex}
%\bibliographystyle{apsrev}

% number  and your conference paper identification number on the
% title page in preprint mode. Multiple \preprint commands are allowed.
%\preprint{BNL--HET--01/BB}

\title{Physics at Future Hadron Colliders}
% Optional argument for running titles on pages
%\title[]{}


% \affiliation command. The \affiliation command should follow th% other information

\author{{\bf Convenors:} U.~Baur (Buffalo), R.~Brock (MSU), J.~Parsons
(Columbia)\\[1.mm]
{\bf Subgroup Convenors:} M.~Albrow (FNAL), D.~Denisov (FNAL),
T.~Han (Wisconsin), A.~Kotwal (Duke), F.~Olness (SMU), J.~Qian
(Michigan)\\[1.mm]
{\bf Working Group:} S.~Belyaev (FSU), M.~Bosman (Barcelona), 
G.~Brooijmans (FNAL),
I.~Gaines (FNAL), S.~Godfrey (Carleton), J.B.~Hansen (CERN), J.~Hauser
(UCLA), U.~Heintz 
(BU), I.~Hinchliffe (LBL), C.~Kao (Oklahoma), G.~Landsberg (Brown),
F.~Maltoni (UIUC), C.~Oleari (Wisconsin), C.~Pagliarone (Pisa), F.~Paige
(BNL), T.~Plehn 
(Wisconsin), D.~Rainwater (FNAL), L.~Reina (FSU), T.~Rizzo (SLAC), S.~Su
(Caltech), T.~Tait (ANL), D.~Wackeroth (Rochester), E.~Vataga (Pisa),
D.~Zeppenfeld (Wisconsin)\\[2.mm]}


\date{\today}

\begin{abstract}
We discuss the physics opportunities and detector challenges at future
hadron colliders. As guidelines for energies and luminosities we use the
proposed luminosity and/or energy upgrade of the LHC (SLHC), and the
Fermilab design of a Very Large Hadron Collider (VLHC). We
illustrate the physics capabilities of future hadron colliders for a
variety of new physics scenarios (supersymmetry, strong electroweak
symmetry breaking, new gauge bosons, compositeness and extra
dimensions). We also investigate the prospects of doing precision Higgs
physics studies at such a machine, and list selected Standard Model
physics rates. 
\end{abstract}
% \pacs{}

%\maketitle must follow title, authors, abstract and \pacs
\maketitle

%%%%%%%%%%%%%%%%%%%%%%%%%%%%%%%%%%%%%%%%%%%%%%%%%%%%%%%%%%%%%%%%%%%%%%%%%%
%%%%%%%%%%%%%%%%%%%%%%%%%%%%%%%%%%%%%%%%%%%%%%%%%%%%%%%%%%%%%%%%%%%%%%%%%%

\section{Introduction}
\label{sec:intro}

Particle physics experiments at the highest possible energies, with
their unique ability to explore the uncharted territory at the energy
frontier, provide exciting opportunities for great advances. For the
past two decades hadron colliders, 
the CERN $p\bar p$ collider and the Fermilab Tevatron, have held the
high energy frontier and have resulted in the discovery of the $W$ and
$Z$-bosons, and the top quark. For at least the next half-decade, the Tevatron
will continue discovery physics experiments at the energy frontier and
in a few years ($\sim 2006$), the LHC will become the
highest energy accelerator by colliding proton beams at a center of mass
energy of $\sqrt{s}=14$~TeV. The LHC will provide excellent chances to
discover the Higgs boson and new physics at the TeV scale. With the
construction of the LHC and its detectors well under way, and in view of
the substantial time needed for planning and construction of large scale
high energy physics projects, it is timely to consider possibilities for
upgrading the LHC and for constructing a new hadron collider 
capable of reaching energies of ${\cal O}(100$~TeV). Such a machine
 should be viewed as post LHC and  Linear Collider (LC)
 and would  allow access to unprecedented
energy scales breaking completely new ground.

For the LHC both a luminosity upgrade by a factor~10 to ${\cal
L}=10^{35}\,{\rm cm}^{-2}\,{\rm s}^{-1}$
%~\cite{ellis} 
and an energy upgrade by a 
factor~2 to $\sqrt{s}=28$~TeV are being discussed (so-called SuperLHC
(SLHC)). Doubling the center
of mass energy of the LHC would require new magnets with a field
strength of about 17~T. Such magnets currently do not exist. In
contrast, a gradual increase of the luminosity by up to a factor~10 by
increasing the bunch intensity to the beam -- beam limit, replacing the
quadrupole magnets near the interaction point, and reducing the bunch
spacing by a factor~2 to 12.5~ns 
appears to be technologically feasible and, if so, can probably be 
achieved within
5~years after the LHC begins operation. From the experimental point of
view, an increase in the center of mass energy is easier to exploit than
an increase in luminosity. Even with major upgrades, the performance of
the LHC detectors ATLAS~\cite{atlas} and CMS~\cite{cms} will be degraded
at ${\cal L}=10^{35}\,{\rm cm}^{-2}\,{\rm 
s}^{-1}$, due to the increased occupancy, radiation and pile-up noise. 
The impact of both
possibilities on the physics capabilities of the ATLAS detector are
discussed in Ref.~\cite{atlasupgr}.

An example of a design study for a Very Large Hadron Collider (VLHC) 
capable of
reaching energies of ${\cal O}(100$~TeV) has recently been presented in
Ref.~\cite{vlhcdesign}. The design  discusses a representative
two  stage $pp$ collider. Both stages  might be housed in a tunnel with
a total circumference 
of 233~km. The first phase (VLHC-I)  would use 2~T super-ferric magnets 
in order to
achieve a total center of mass energy of $\sqrt{s}=40$~TeV and a
luminosity of ${\cal L}=10^{34}\,{\rm cm}^{-2}\,{\rm s}^{-1}$. If the
technology of super-ferric magnets can be pushed to 3~T as
anticipated~\cite{m4talk}, the center of mass energy of the VLHC-I  could
be increased to about 50~TeV. Stage~2 of the VLHC (VLHC-II)  would
make use of the VLHC-I
ring as an injector and  could reach energies between 125~TeV and 200~TeV,
depending on the field strength of the magnets~\cite{vlhcdesign}. The
 luminosity for $\sqrt{s}=125$~TeV would be of order ${\cal L}=5.1\cdot 
10^{34}\,{\rm cm}^{-2}\,{\rm s}^{-1}$, and would gradually  decrease to 
approximately ${\cal L}=2.1\cdot 10^{34}\,{\rm cm}^{-2}\,{\rm s}^{-1}$ at
$\sqrt{s}=200$~TeV. At  VLHC-I, the number of interactions per bunch
crossing  would be similar to that at the LHC ($\approx 20$). At 
VLHC-II, one would expect  between~50 and~100 interactions per bunch crossing,
depending of the center of mass energy. Luminosities up to ${\cal
L}=10^{35}\,{\rm cm}^{-2}\,{\rm s}^{-1}$ at the highest center of mass 
energies considered are possible if the
heat generated by synchrotron radiation~\cite{m4talk} can be efficiently
removed. Furthermore, the design can be modified to a single stage
machine with $\sqrt{s}=150-200$~TeV. In that  scenario, a separate 5~TeV
injector  would be built in a new 15~km circumference tunnel, which could also
house a GigaZ $e^+e^-$ collider~\cite{m4talk}. 

In order to reach a center of mass energy of 200~TeV in the VLHC tunnel,
superconducting high field magnets with a field strength of about 11~T
would be necessary. Such magnets could in principle also be installed in the
Tevatron tunnel, or a site---filling tunnel at Fermilab, raising the
possibility of a $p\bar p$ collider with a luminosity of 
${\cal L}=10^{33}\,{\rm cm}^{-2}\,{\rm s}^{-1}$  and a center of mass
energy of $\sqrt{s}=5.4$~TeV and 12~TeV,
respectively. Studies have shown~\cite{tripler} that such a
machine has a discovery potential for the Higgs boson and supersymmetry
which is similar to that of the LHC. However, it would be very difficult to
complete such a project before 2010, approximately 4~years after the LHC
will commence operation.

\begin{figure}
\includegraphics{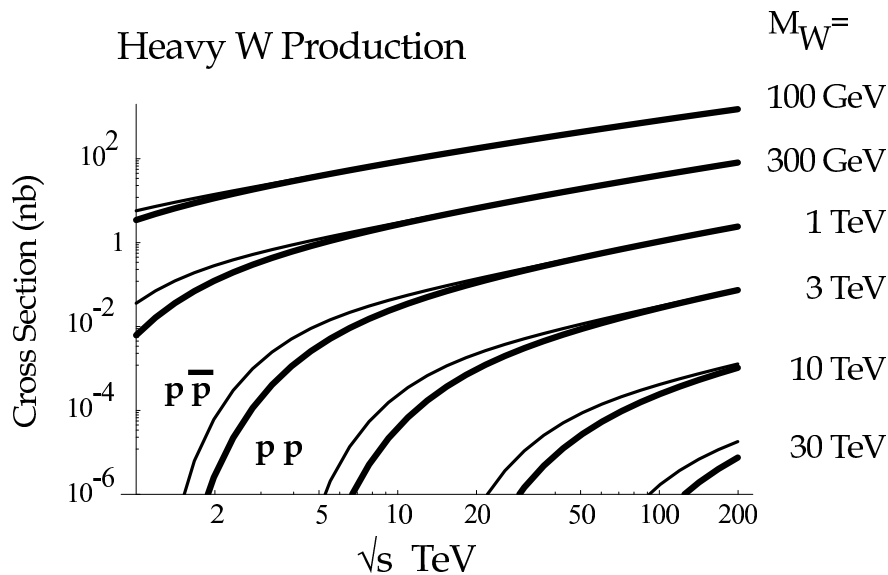}%
\caption{The total cross section for producing a heavy $W$ boson with SM
like couplings in $pp$ and $p\bar p$ collisions as a function of the
center of mass energy. For each value of $M_W$, the upper (thin) line
gives the cross section for $p\bar p\to W^\pm$, while the lower
(thick) line shows the cross section $pp\to W^\pm$.}
\label{fig:one}
\end{figure}

With the Tevatron now running and the LHC on the horizon, it is
appropriate to think ahead --  to begin the investigation of  the
physics potential of and the 
detector requirements for hadron
colliders with center of mass energies reaching the 100~TeV region and/or
luminosities which are up to a factor~10 larger than the LHC design
luminosity. In this report we explore both aspects for a
generic very large hadron collider (VLHC), selecting 
\begin{itemize}
\item $\sqrt{s}=40$~TeV (VLHC-I),
\item $\sqrt{s}=100$~TeV (VLHC-II, low energy), and
\item $\sqrt{s}=200$~TeV (VLHC-II, high energy)
\end{itemize}
as reference energies. Results will be presented for integrated
luminosities in the range of 100~fb$^{-1}$ to 1~ab$^{-1}$. Whenever
possible, we compare the physics reach of a VLHC with that of the
SLHC. We shall assume that the VLHC is a $pp$ collider. For a given
center of mass energy, the production cross section for a particle of
mass $M$ in $pp$ and $p\bar p$ collisions is very similar due to the
large sea -- sea quark flux. The valence quark -- anti-quark flux only
contributes measurably to the cross section for $M>0.2\,\sqrt{s}$. This
is illustrated in Fig.~\ref{fig:one} for a heavy $W$ boson with Standard
Model (SM) like couplings. Since the luminosity of a $p\bar p$ collider
is estimated to be about a factor~10 smaller than that of a $pp$
collider with the same center of mass energy, we shall only consider
$pp$ collisions in the following.

The remainder of this report is organized as follows. In
Section~\ref{sec:two} we consider in some detail the physics
opportunities at a VLHC. We discuss precision Higgs boson physics, the
search for supersymmetry, and aspects of strong electroweak
symmetry breaking relevant for future hadron colliders. We also discuss
the search reach for new gauge bosons, compositeness of quarks and
leptons, and extra dimensions. For completeness, we also give
results for selected SM cross sections. At the LHC and a linear $e^+e^-$
collider, the search for the Higgs boson and precision Higgs boson
physics provide well-motivated, specific ``physics cases''. Not
surprisingly, this is not the case for a VLHC. Since such a machine will
explore {\sl terra incognita}, we can speculate about the physics to be
discovered with such a facility, but cannot at present make the case 
that a particular
discovery is likely to occur. In Section~\ref{sec:three} we
discuss the detector requirements for a VLHC and outline a R\&D
program necessary in order to achieve the detector performance needed. 
Finally, in
Section~\ref{sec:four}, we summarize our results and discuss future
perspectives. 

\section{Physics Potential of Future Hadron Colliders}
\label{sec:two}

In this Section, we explore the physics opportunities at a VLHC. All
results reported here are preliminary and represent only first steps
towards a more complete analysis of the physics capabilities of such a
machine. As noted above, a VLHC will come after at least two discovery
machines (Tevatron and LHC) 
will have had considerable experience. We expect that the physics
panorama will be different from what we 
face today. Therefore, it is important to note that
the various scenarios considered often serve as place-holders for the
physics menu of 15 -- 20 
years from now. We can be confident that pathways to this new physics
will still pass through the need to be able to  measure jets, leptons,
photons and missing energy and have to function within radiation and rate
environments which can be reliably predicted. However, the physics
reactions chosen primarily act as illustrations of capabilities and 
indications of challenges which might be faced. They do not pretend to 
function as 2001 justifications for a 20?? VLHC or SLHC machine. The 
lead-time for design and construction will possibly prevent us from
working within a realm of confident prediction of specific physics 
predictions. Whenever appropriate we compare the physics reach of the VLHC
with that of the SLHC and that of linear $e^+e^-$ colliders under
consideration, such as Tesla~\cite{tesla}, the NLC~\cite{resource},
the JLC~\cite{acfa} ($\sqrt{s}=500-1500$~GeV, ${\cal
L}={\rm few}\times 10^{34}\,{\rm cm}^{-2}\,{\rm 
s}^{-1}$), or CLIC $(\sqrt{s}=3-5$~TeV, ${\cal
L}=10^{35}\,{\rm cm}^{-2}\,{\rm s}^{-1})$~\cite{clicdesign,marco}. 

\subsection{Standard Model Cross Sections}

By the time a VLHC-I would begin  operation  those parameters of the SM which
are currently not well constrained, such as the top quark mass, $m_t$, the
Higgs boson mass, $M_H$, or the gauge boson self-interactions, should be known
very accurately from data collected at the LHC and/or a linear $e^+e^-$
collider. The reason for building a VLHC therefore is {\sl not} to test
the SM but to directly search for new physics. However, many SM
processes, such as top quark production, weak boson or weak boson pair
production, are important sources of backgrounds for new physics
processes. Accurate knowledge of the cross sections for SM processes,
including higher order QCD and electroweak corrections  is
necessary. 

\begin{figure}
\begin{tabular}{cc}
\includegraphics[width=3.25in]{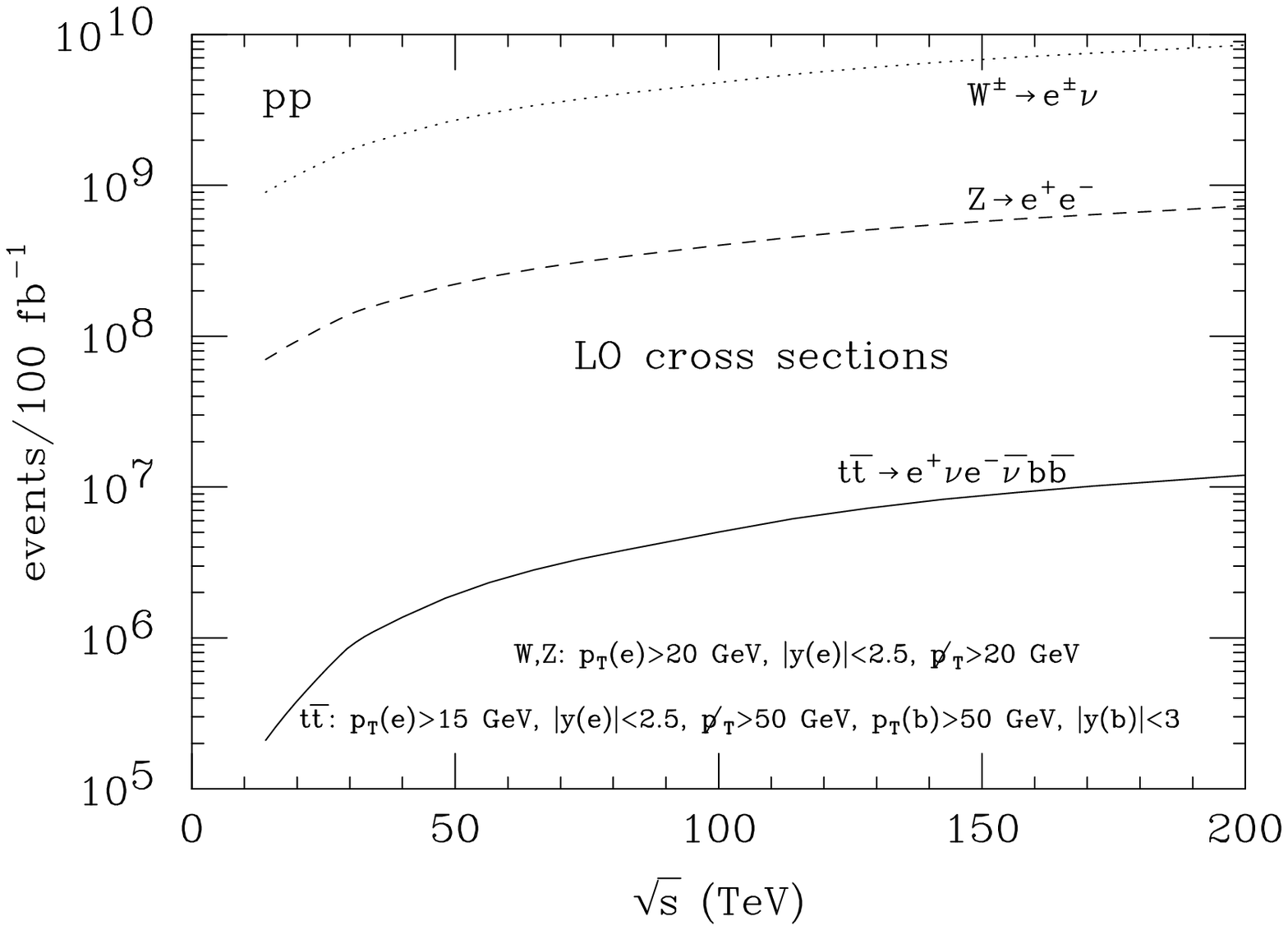} &
\includegraphics[width=3.25in]{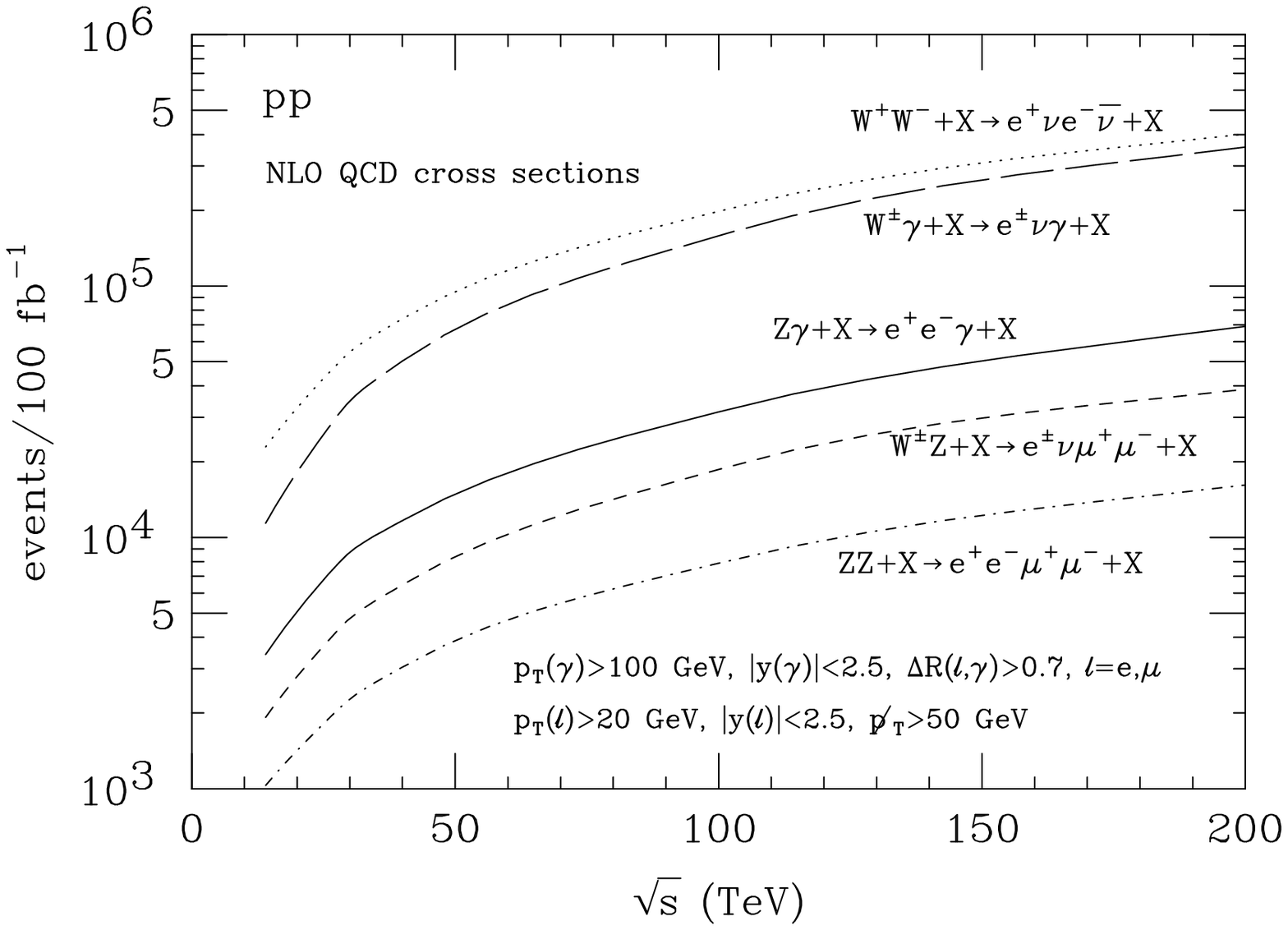} 
\end{tabular}
\caption{Event rates in $pp$ collisions for selected SM processes as a
function of $\sqrt{s}$. The cuts imposed are shown in the figure.}
\label{fig:two}
\end{figure}

In Fig.~\ref{fig:two} we show expected event rates for an
integrated luminosity of 100~fb$^{-1}$ for selected SM processes as a
function of $\sqrt{s}$, taking into account minimal acceptance cuts 
listed on the figures for the observed final state particles.
The cross sections of SM processes grow by about a factor 10~--~50 as
$\sqrt{s}$ is increased from 14~TeV to 200~TeV. 
The left panel shows that $W\to e\nu$ and $\bar tt\to e^+\nu e^-\bar\nu\bar
bb$ events are produced with a rate of about 1~kHz and 1~Hz respectively
 for $\sqrt{s}=200$~TeV. This may allow for indirect searches of new physics
in rare decays, such as $t\to WZb$ (SM: BR($t\to WZb)\approx 2\cdot
10^{-6}$~\cite{alta}). The right panel displays the diboson cross
sections at NLO in QCD. QCD corrections enhance the diboson rates by a
factor 1.3~--~2.8 at LHC energies~\cite{dibos}, and by a factor
1.9~--~5.0 at $\sqrt{s}=200$~TeV. The size of the QCD corrections is
smallest for $Z\gamma$ production, and largest for $W\gamma$
production. The large enhancement of the diboson cross sections is due
to a logarithmic enhancement factor in the $qg$ and $\bar qg$ real emission
subprocesses. The cross section for $ZZ$ production also
includes the contribution from gluon fusion~\cite{ggzz}. At LHC and VLHC
energies, the $gg\to ZZ$ cross section is not negligible compared to the
lowest order $\bar qq\to ZZ$ rate:
\begin{eqnarray}
\sigma(gg\to ZZ)\approx 0.13\,\sigma(\bar qq\to ZZ) & {\rm for} &
\sqrt{s}=14~{\rm TeV,}\\
\sigma(gg\to ZZ)\approx 0.45\,\sigma(\bar qq\to ZZ) & {\rm for} &
\sqrt{s}=200~{\rm TeV.}
\end{eqnarray}
The cross section for $gg\to W^+W^-$~\cite{kao} is less than 10\% of the
lowest order $\bar qq\to W^+W^-$ cross section over the entire
$\sqrt{s}$ considered here, and is therefore not included in
Fig.~\ref{fig:two}. 

\subsection{Precision Higgs Boson Physics}

If the SM Higgs boson exists, it will be discovered either at the
Tevatron~\cite{tevhiggs} or the LHC~\cite{atlastdr}. The Tevatron will
be able to find the Higgs boson if its mass is less than about 170~GeV, 
assuming an integrated luminosity of 30~fb$^{-1}$ can be achieved per
experiment. In contrast, the LHC can find the SM Higgs
boson~\cite{atlastdr} over the 
entire range from the present lower experimental limit of
$M_H>114.1$~GeV~\cite{higgslim} to the TeV region. From a global fit to
the electroweak observables, one obtains a 95\% CL upper limit of $M_H<196$~GeV
from present data~\cite{charlton}. In the following we concentrate on a
SM Higgs boson with mass in the range $114~{\rm GeV}<M_H<200$~GeV. The
discovery potential of Higgs bosons in the minimal supersymmetric SM at
the LHC and VLHC is discussed in Refs.~\cite{highpt,cmsmssm} and~\cite{ckao},
respectively. 

\subsubsection{Measurement of Higgs boson properties at the LHC/SLHC and
a Linear Collider}
\label{sec:lhc}

Once the Higgs boson has been
discovered, it will be important to measure its properties in order to
test whether its coupling to fermions and gauge bosons is as predicted
by the SM. Even more important will be the measurement of the Higgs
boson self-coupling, $\lambda_{HHH}$, which partly determines the shape of the
Higgs potential. In the SM, $\lambda_{HHH}=3M_H^2/2v$, where $v=246$~GeV
is the vacuum expectation value. An accurate test of this relation may
reveal whether the minimal Higgs sector of the SM, or an extended sector
which arises for example in supersymmetric theories, is realized. 

With an integrated
luminosity of 300~fb$^{-1}$, ATLAS and CMS combined are expected to 
measure $M_H$ with a precision
of about 0.1\%~\cite{highpt,hohl}. One also hopes to determine the total
width of the Higgs boson, $\Gamma_H$, 
and $\sigma\times$Br for $H\to\gamma\gamma$, $H\to ZZ^{(*)}\to
4$~leptons and $H\to WW^{(*)}\to\ell\nu\ell\nu$ with a precision of
about 10\%~\cite{hohl}, ratios of couplings to gauge bosons and fermions
with an accuracy of $10-20\%$~\cite{ritva}, and the $t\bar tH$ coupling,
$y_{ttH}$, with a precision of 13\%~\cite{drollinger}. Since many of these
measurements at the LHC are statistically limited, one can hope to improve
the precision which can be achieved by a factor~1.5 to~2 if the LHC
energy could be doubled to $\sqrt{s}=28$~TeV. As discussed in
Sec.~\ref{sec:tagging}, jet tagging and central jet veto, which are
crucial for determining the Higgs boson couplings at the
LHC~\cite{ritva}, become less efficient if the luminosity is increased
beyond ${\cal L}=10^{34}\,{\rm cm}^{-2}\,{\rm s}^{-1}$. The measurement
of the Higgs boson couplings is therefore expected to profit only
modestly (by {\sl at most} a factor~2)
from a luminosity upgrade of the LHC~\cite{ellis}. If $115~{\rm
GeV}<M_H<140$~GeV, 
the $H\mu\mu$ coupling may be determined with a precision of about 15\%
at the LHC, assuming that the $t\bar tH$ and $b\bar bH$ couplings are
SM-like~\cite{han}. In order to probe 
the Higgs boson self-coupling, Higgs pair production processes have to
be studied. The
cross sections of the $H$ pair signal processes at the LHC have been
calculated in Ref.~\cite{maggie}, however, a study of the feasibility 
of measuring $\lambda_{HHH}$ has not yet
been carried out. Such a measurement may well be very difficult at the LHC.

While the LHC will make it possible to perform first, but still somewhat
rough, measurements of the Higgs boson properties, a LC will provide the
opportunity to carry out more precise measurements of the couplings of
the Higgs boson to gauge bosons and the fermions which belong to the
third generation ($\tau$, $b$ and $t$)~\cite{comp}. 
At a LC with $\sqrt{s}=500$~GeV and 500~fb$^{-1}$, the couplings of the
Higgs boson to $b$-quarks, $\tau$-leptons, 
photons, $W$-bosons and gluons can be measured with a precision of a few
per cent~\cite{tesla,resource,acfa}. In addition, the top quark Yukawa
coupling $y_{ttH}$ can be 
determined with a precision of about 5.5\% (21\%) at $\sqrt{s}=800$~GeV
($\sqrt{s}=500$~GeV) and with 1~ab$^{-1}$. A LC will also make it
possible to perform a
measurement of the $H\mu\mu$ and Higgs boson self-coupling, although a
multi-TeV collider with a luminosity of ${\cal
L}=10^{35}\,{\rm cm}^{-2}\,{\rm s}^{-1}$ or more is required for a 
precise determination of these parameters. 
Recent studies have shown that the 
$H\mu\mu$ coupling can be determined with a precision of about 15\% (4\%)
for $M_H=120$~GeV and an integrated luminosity of 1~ab$^{-1}$
(5~ab$^{-1}$) at a LC 
operating at $\sqrt{s}=800$~GeV ($\sqrt{s}=3$~TeV)~\cite{albert}, and
that the Higgs boson self-coupling can be measured with an accuracy of
about 20\% ($7-8\%$) for $\sqrt{s}=500$~GeV ($\sqrt{s}=3$~TeV) and
5~ab$^{-1}$~\cite{tesla,resource,krakow}. The most natural place for
extracting the muon Yukawa coupling of course is a muon collider~\cite{raja},
operating at energies around the Higgs boson mass. Assuming that the
branching ratio for $H\to\bar bb$ can be determined to 2.5\% at a LC, it
should be possible to measure the $H\mu\mu$ coupling at such a machine
with a precision of about 2\%~\cite{vernon}. 

\subsubsection{Jet Tagging at VLHC Energies}
\label{sec:tagging1}

\begin{figure}
\includegraphics[width=3.5in]{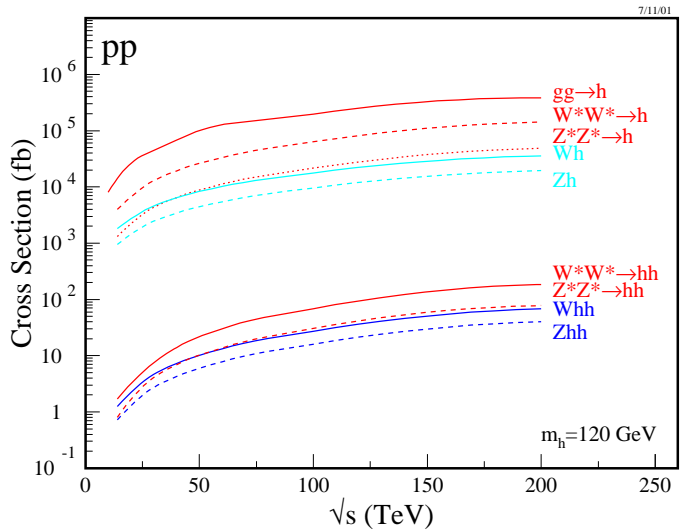}%
\caption{Cross sections for various Higgs and Higgs pair production
processes as a function of $\sqrt{s}$ for $pp$ collisions and 
$M_H=120$~GeV.}
\label{fig:three}
\end{figure}

\begin{figure}
\begin{tabular}{cc}
\includegraphics[width=3.25in]{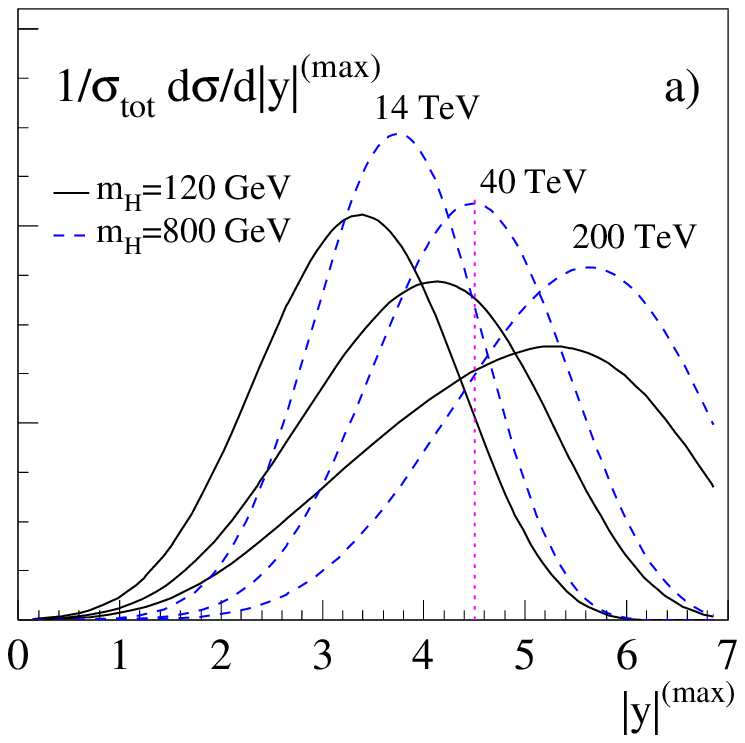} &
\includegraphics[width=3.25in]{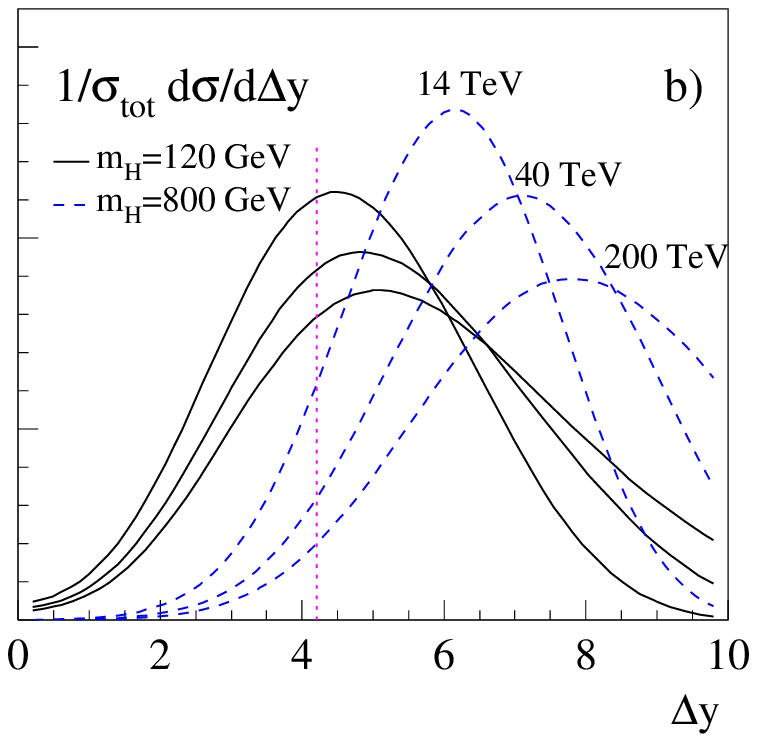} 
\end{tabular}
\caption{a) Normalized maximum jet rapidity distribution and b)
normalized rapidity difference distribution for the WBF signal for two
values of $M_H$ and several center of mass energies. A
$p_T(j_{1,2})>30$~GeV cut is imposed on the tagging jets. The figure is
taken from Ref.~\cite{plehn}.}
\label{fig:four}
\end{figure}

It is interesting to investigate whether an upgraded LHC or a VLHC 
offer a chance to improve the LC measurements of the Higgs boson
couplings, in particular those of the $H\mu\mu$, the $t\bar
tH$, and the $HHH$ couplings. The cross sections for Higgs production
processes increase by about a factor $10-30$ from the LHC energy to the
highest VLHC-II energy (see Fig.~\ref{fig:three}).
In order to obtain information on the couplings of the Higgs boson at
hadron colliders, it is crucial to make use of Higgs boson production
via weak boson fusion (WBF), i.e. by separately observing $qq\to qqH$
and crossing related processes in which the Higgs boson is radiated off
a $t$-channel $W$ or $Z$~\cite{ritva}. WBF events are characterized by
two forward jets which are separated by a large rapidity gap. Forward
jet tagging and a central jet veto thus are crucial to reduce the QCD
background and to extract the WBF signal. The characteristics of the
tagging jets in WBF at the LHC and at VLHC energies are shown in
Fig.~\ref{fig:four} for $M_H=120$~GeV and $M_H=800$~GeV. 
It should be
noted that very similar results are obtained if the Higgs boson is
replaced by another massive (possibly composite) 
particle of equal mass as long as it couples
to $W$ and $Z$ bosons and is produced via WBF. Such states frequently occur
in models of dynamical symmetry breaking~\cite{sekhar}. 
Figure~\ref{fig:four}a  shows the normalized
differential cross section as a function of
$|y|^{max}=\max(|y(j_1)|,|y(j_2)|)$, where $y(j_1)$ and $y(j_2)$ are the
rapidities of the tagging jets. Figure~\ref{fig:four}b  displays the 
normalized differential cross section as a function of the rapidity
difference of the two jets, $\Delta y=|y(j_1) - y(j_2)|$. 
At the LHC, the $|y|^{max}$ distribution peaks at $|y|^{max}\approx
3-3.5$. With 
increasing machine energy, the peak moves to higher values of rapidities
and broadens. For $\sqrt{s}=200$~TeV, the maximum occurs at
$|y|^{max}\approx 5.5$ with a long tail extending to values larger
than~7. The $\Delta y$ distribution peaks at $\Delta y\approx 4-4.5$ for
a light Higgs boson for the energies considered. For a heavy state produced
via WBF, the rapidity gap between the two jets is more pronounced and
widens considerably with higher values of $\sqrt{s}$. For a 200~TeV
collider, the tagging jets are separated by typically 8 -- 10 units in
rapidity for production of a heavy Higgs boson. 

From Fig.~\ref{fig:four}  one concludes that it will be necessary to
have a hadron calorimeter which covers the region out to $|y|=6-7$
in order to take full advantage of jet tagging in WBF. The vertical
dotted line in Fig.~\ref{fig:four}a 
indicates the maximum rapidity for
which the LHC detectors are efficient in detecting forward jets. If this
cannot be improved, a large fraction of the WBF signal will be lost at
the VLHC. 

\subsubsection{$H\to\mu^+\mu^-$ and $H\to\tau^+\tau^-$ at the VLHC}

The search for Higgs boson decays to muons is rate limited due to
the extremely small $H\mu\mu$ coupling. For $gg\to
H\to\mu^+\mu^-$, the processes $q\bar q\to\gamma, Z\to\mu^+\mu^-$ 
constitute a large
irreducible background. In WBF, on the other hand, jet tagging and
central jet veto offer a handle to suppress the QCD $\mu^+\mu^- jjX$
background. The significance of the $H\to\mu^+\mu^-$ signal in WBF at the
LHC and the VLHC for several values of $M_H$ are shown in
Table~\ref{tab:one}~\cite{plehn}. 
\begin{table}
\caption{Significance of the $H\to\mu^+\mu^-$ signal in WBF for
$\int\!{\cal L}dt=300~{\rm fb}^{-1}$ at the LHC and VLHC (from
Ref.~\cite{plehn}).}
\label{tab:one}
\begin{tabular}{|c|c|c|c|}
\tableline
$M_H$ (GeV) & $\sqrt{s}=14$~TeV (LHC) & $\sqrt{s}=40$~TeV (VLHC-I) & 
$\sqrt{s}=200$~TeV (VLHC-II) \\
\tableline
120 & 1.8 & 3.3 & 5.7 \\
130 & 1.7 & 3.2 & 5.3 \\
140 & 1.2 & 2.4 & 4.0 \\
\tableline
\end{tabular}
\end{table}
The following WBF selection cuts were imposed in Table~\ref{tab:one}:
\begin{equation}
p_T(j_{1,2})> 20~{\rm GeV}, \quad\Delta R(j_1,j_2)>0.6,\quad
|y(j_{1,2})|<4.5, \quad |y(j_1)-y(j_2)|>4.2, \quad y(j_1)\cdot y(j_2)<0.
\label{eq:hcuts}
\end{equation}
In order to reduce the $t\bar t$, $b\bar b$ and $Zjj$ backgrounds, an invariant
mass cut on the two jet system was imposed:
\begin{equation}
m(jj)> 500~{\rm GeV}\quad {\rm (LHC)} \quad {\rm and} \quad 
m(jj)> 1000~{\rm GeV} \quad {\rm (VLHC).} 
\label{eq:hcuts1}
\end{equation}
At VLHC-II, the decay $H\to\mu^+\mu^-$ should be detectable for the
Higgs boson masses studied, and, for an integrated luminosity of
1~ab$^{-1}$, it should be possible to measure the $H\mu\mu$ coupling
with a precision similar to that which can be achieved at CLIC with
5~ab$^{-1}$ (assuming that the coupling of the Higgs boson to the $W$
and $Z$ bosons has been precisely measured before, eg. at a LC). If one
includes the $gg\to H\to\mu^+\mu^-$ channel, the precision improves by
about a factor~1.5~\cite{han}, provided that the $Hgg$ coupling and
higher order QCD corrections are
known with sufficient precision. Further strengthening of the signal in
WBF is possible if one assumes that the hadron calorimeter coverage 
extends to rapidities of $|y(j_{1,2})|\approx 6-7$ instead of the value
listed in~(\ref{eq:hcuts}). 

In order to see whether the measurement of the Higgs boson couplings to 
the third generation fermions at the LC can be improved at a VLHC, the
$H\to\tau^+\tau^-$ channel in WBF was analyzed~\cite{zep}. Imposing
similar jet tagging cuts as in the $H\to\mu^+\mu^-$ case
(see~(\ref{eq:hcuts}) and ~(\ref{eq:hcuts1})) it was found that, for
$M_H=120-140$~GeV, the $H\tau\tau$ coupling can be measured at a VLHC
with a statistical uncertainty of a few per cent for an integrated
luminosity of 100~fb$^{-1}$. This is comparable with the precision
expected at a LC. However, for $\sqrt{s}>100$~TeV, gluon
fusion becomes a pronounced source of $H+2$~jet events~\cite{oleari}, and
the clean separation of gluon fusion and WBF cross sections will require
additional effort. More detailed studies are also necessary in order to
assess the experimental systematic uncertainties. 

\subsubsection{Measuring the Top Quark Yukawa Coupling at the VLHC in
$t\bar tH$ Production}

\begin{figure}
   \begin{minipage}[b]{.46\linewidth}
    \hspace{-0.7truecm}
    \includegraphics[height=3.5in]{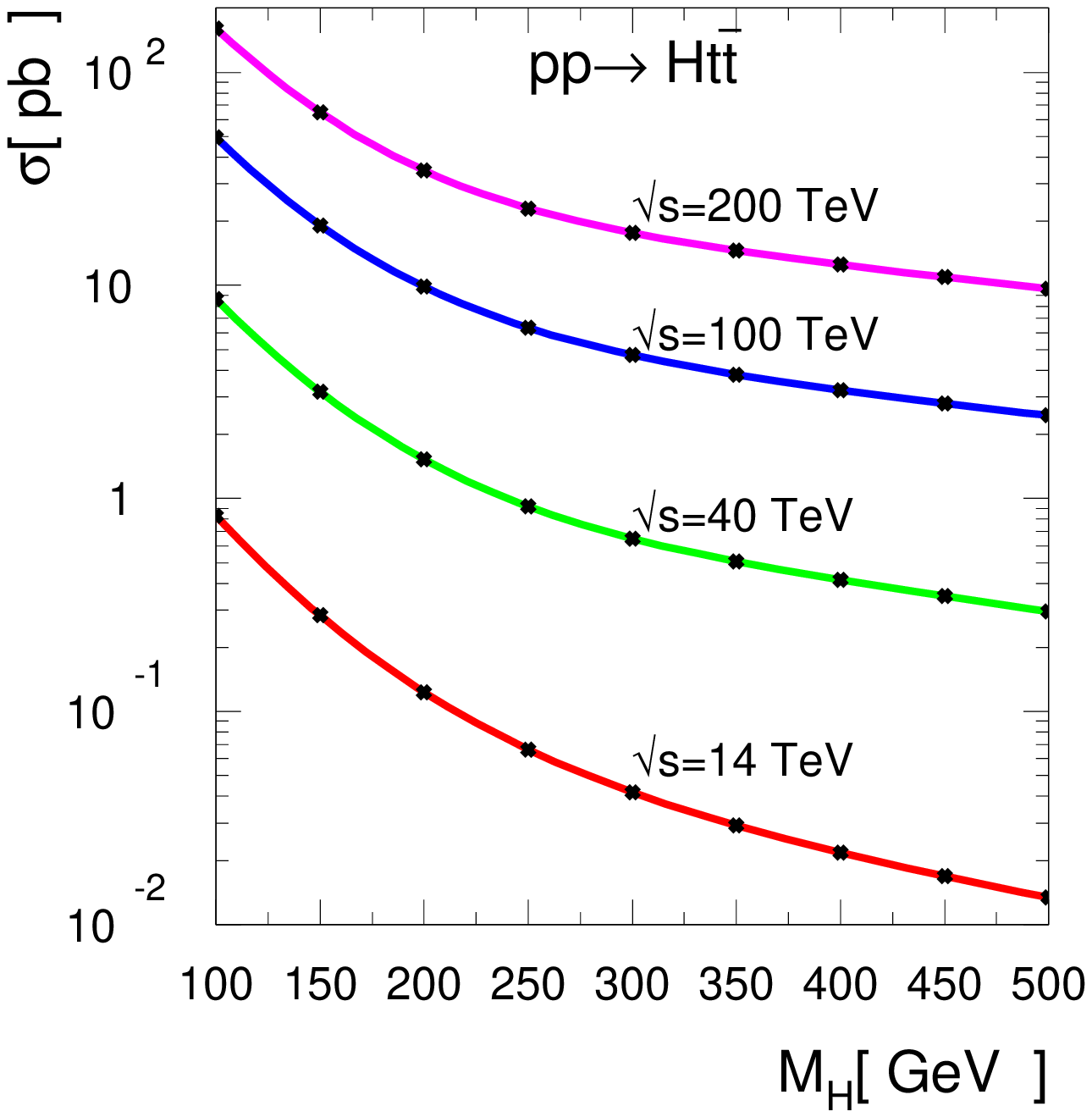}
    \vspace{-0.2cm}
    \caption{The leading order cross section for $pp\rightarrow t\bar tH$
      as a function of $M_H$ for various center of mass
      energies in the SM. The CTEQ4L set of parton distribution
functions has been 
used to evaluate the cross section, and the renormalization and
factorization scales have been set to $\mu=m_t+M_H/2$ with $m_t=174$~GeV. }
    \label{fig:five}
%    \vspace{-0.4truecm}
  \end{minipage}
   \hspace{0.2truecm}
   \begin{minipage}[b]{.46\linewidth}
    \hspace{0.5truecm}
    \includegraphics[scale=0.413]{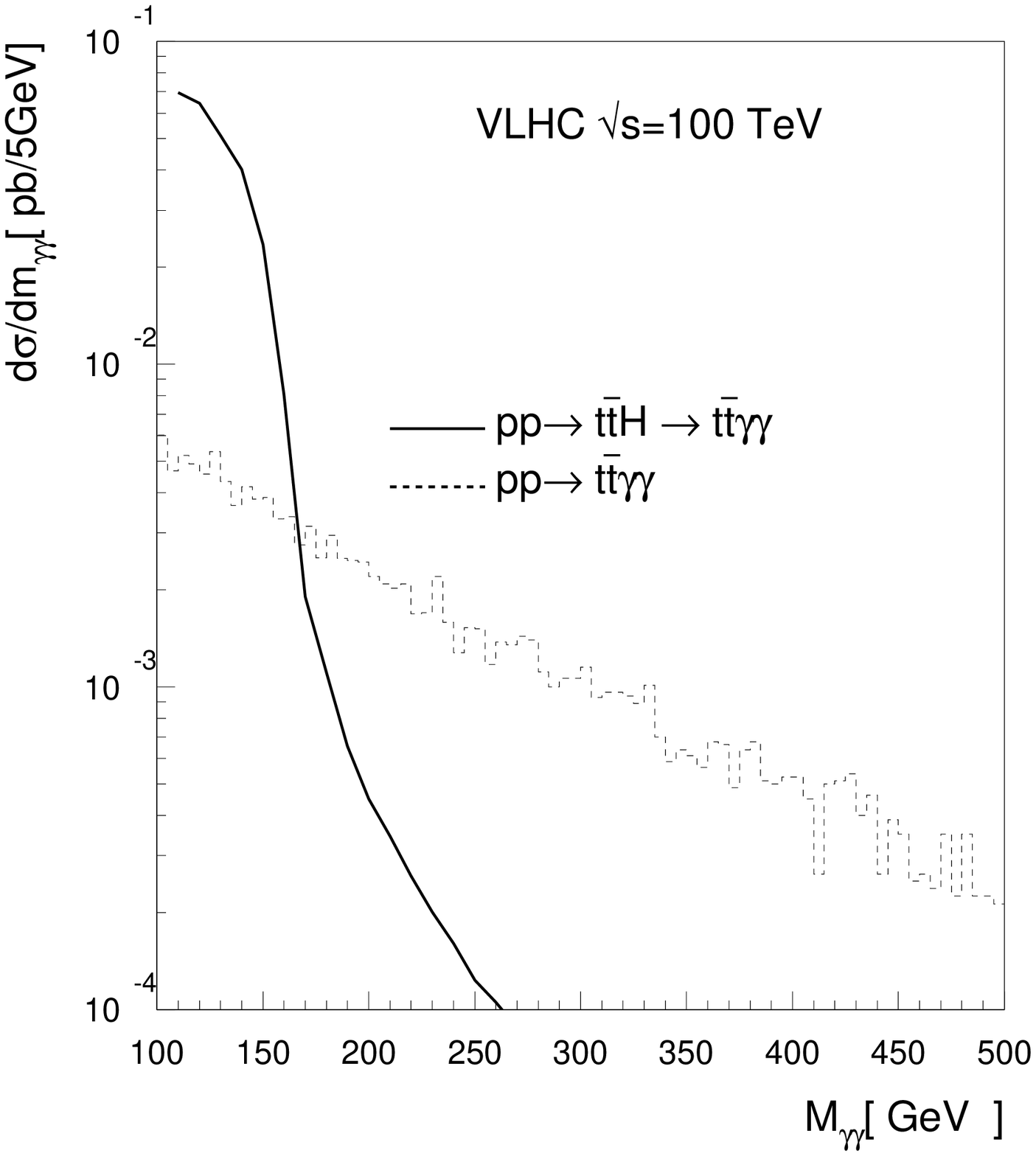}
    \vspace{0.3cm}
    \caption{Cross sections for the SM $\bar ttH(\to\gamma\gamma)$ signal 
(solid line) and the irreducible $t\bar t\gamma\gamma$ 
background  (dashed histogram) at 
$\sqrt{s}=100$~TeV. To simulate detector acceptance we require
  $p_T(\gamma)>25$ GeV and $|\eta(\gamma)|<3$.}
    \label{fig:six}
    \vspace{0.7truecm}
  \end{minipage}
\end{figure}

The top quark Yukawa coupling at a hadron collider is most easily
measured using $t\bar tH$ production. At the highest VLHC-II energies,
the SM 
$t\bar tH$ cross section is a factor $100-1000$ larger than at the
LHC~\cite{fabio} (see Fig.~\ref{fig:five}).
The $t\bar tH$ production cross section at VLHC-II energies is
sufficiently large that even rare decays such as $H\to\gamma\gamma$ can
be observed in this mode. The SM $t\bar tH(\to\gamma\gamma)$ signal and the
$t\bar t\gamma\gamma$ continuum background for $pp$ collisions at
$\sqrt{s}=100$~TeV are shown in Fig.~\ref{fig:six}. For $M_H<150$~GeV,
the background is seen to be totally negligible. Taking into account
realistic particle detection efficiencies and assuming that the
$H\gamma\gamma$ coupling has been measured at the LHC or a linear
collider, the top Yukawa coupling can be measured with an accuracy of
about 7\% for $M_H=130$~GeV and $\sqrt{s}=100$~TeV~\cite{fabio}. A
similar analysis for $H\to\tau^+\tau^-$ and $H\to\bar bb$ shows that
$y_{ttH}$ can be determined with a precision of 3.5\% and 1.5\% in these
channels. Of course this assumes that the $H\tau\tau$ and $H\bar bb$
couplings have been determined before, e.g. at the LHC ($H\tau\tau$) or
the LC ($H\tau\tau$, $H\bar bb$). Alternatively, from the ratio of the
cross sections in the $t\bar t\tau^+\tau^-$ and $t\bar tb\bar b$
channels one can measure the strength of the $H\bar bb$ coupling. For
$\sqrt{s}=200$~TeV, the precision which one hopes to achieve for
$y_{ttH}$ improves by a factor~1.5 to~2 over that obtained for
$\sqrt{s}=100$~TeV.

\begin{figure}
\includegraphics[width=4in]{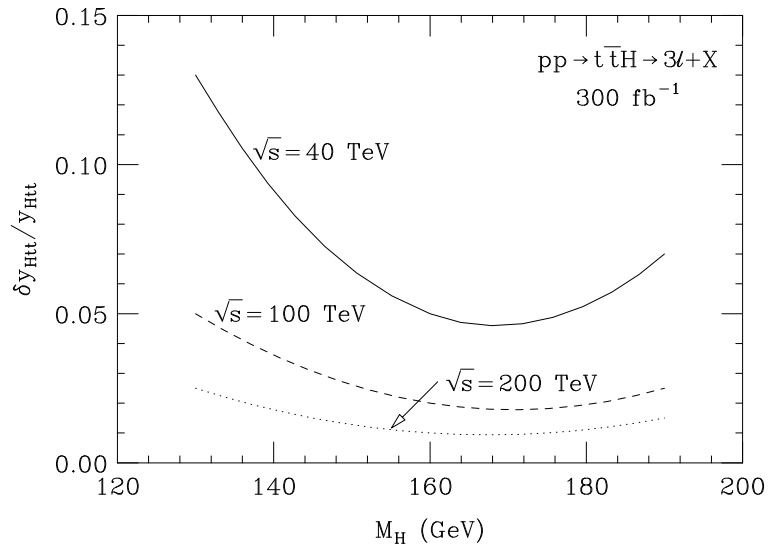}%
\caption{The relative precision of the top quark Yukawa coupling as a
function of the mass of the Higgs boson which can be obtained for an
integrated luminosity of 300~fb$^{-1}$ in $pp\to
\to t\bar tH(\to W^+W^-)\to 3\ell+X$. To simulate detector response, the
following cuts were imposed: $p_T(j)>30$~GeV, $p_T(b)>30$~GeV,
$|\eta(j)|<4.5$, $|\eta(b)|<2.5$, $p_T(\ell)>15$~GeV, 
$|\eta(\ell)|<2.5$, and $p\llap/_T>50$~GeV. In addition, events with 
$81~{\rm GeV}<m(\ell^+\ell^-)<101$~GeV are excluded. $b$-quarks
(leptons) are assumed to be detected with an efficiency of 50\% (85\%).}
\label{fig:seven}
\end{figure}

If the Higgs boson is in the mass range $140~{\rm GeV}<M_H<190$~GeV,
important information on the top quark Yukawa coupling can also be
obtained from $pp\to t\bar tH(\to W^+W^-)$. One finds that the highest
sensitivities are obtained from the $3\ell+X$ final
state. Figure~\ref{fig:seven} shows the precision for $y_{ttH}$ which
one expects to achieve for 300~fb$^{-1}$ as a function of $M_H$ and
several center of mass energies. The backgrounds from 
$t\bar t ZX$, $t\bar t WX$, $t\bar t WW$ and $t\bar tt\bar t$ production
are taken into account in this analysis~\cite{rain2}.
At the VLHC-I (VLHC-II), a measurement of $\delta y_{ttH}/y_{ttH}\approx 0.05 -
0.10$ ($\delta y_{ttH}/y_{ttH}\approx 0.01 - 0.02$) should be possible.

A VLHC will thus be able to considerably improve the measurement of the
top quark Yukawa coupling over the precision which can be achieved at a
linear $e^+e^-$ collider (see Sec.~\ref{sec:lhc}).

\subsubsection{Can one probe the Higgs Potential at the VLHC?}
\label{sec:2higgs}

As mentioned before, Higgs pair production offers an opportunity to
probe the Higgs boson self-coupling, $\lambda_{HHH}$. As can be seen
from Fig.~\ref{fig:three}, the cross sections for Higgs pair production
processes are a factor~100 to~1000 smaller than those for single Higgs
production. Taking into account the small branching fractions for final
states with manageable background, it is clear that it will be very
difficult to measure $\lambda_{HHH}$ in a hadron collider
environment. The most promising decay channel may be $HH\to\bar
bbW^+W^-\to \bar bb\ell_1\ell_2p\llap/_T$. The branching ratio for this
channel peaks at $\approx 0.8\%$ for $M_H\approx 130$~GeV. 
No detailed study of Higgs pair production at hadron
colliders has been carried out so far. For $gg\to HH$, backgrounds from
QCD induced processes are likely to be severe. Excellent $b$-tagging 
may help to reduce the background to an acceptable level. 
On the other hand, as in single
Higgs boson production, the characteristics of WBF should make it possible
to reduce the background to an acceptable level in $qq\to qqHH$. 
A simple estimate then shows that a minimum of several ab$^{-1}$ is needed
in order to be able to
measure $\lambda_{HHH}$ in this channel with a precision similar to that
which has been predicted for a multi-TeV $e^+e^-$
collider~\cite{krakow}. 

\subsubsection{Jet tagging and Central Jet Veto at ${\cal
L}=10^{35}\, cm^{-2}\, s^{-1}$}
\label{sec:tagging}

\begin{figure}
   \begin{minipage}[b]{.46\linewidth}
    \hspace{-0.7truecm}
\includegraphics[height=3.3in,clip,bb=190 500 430 
720]{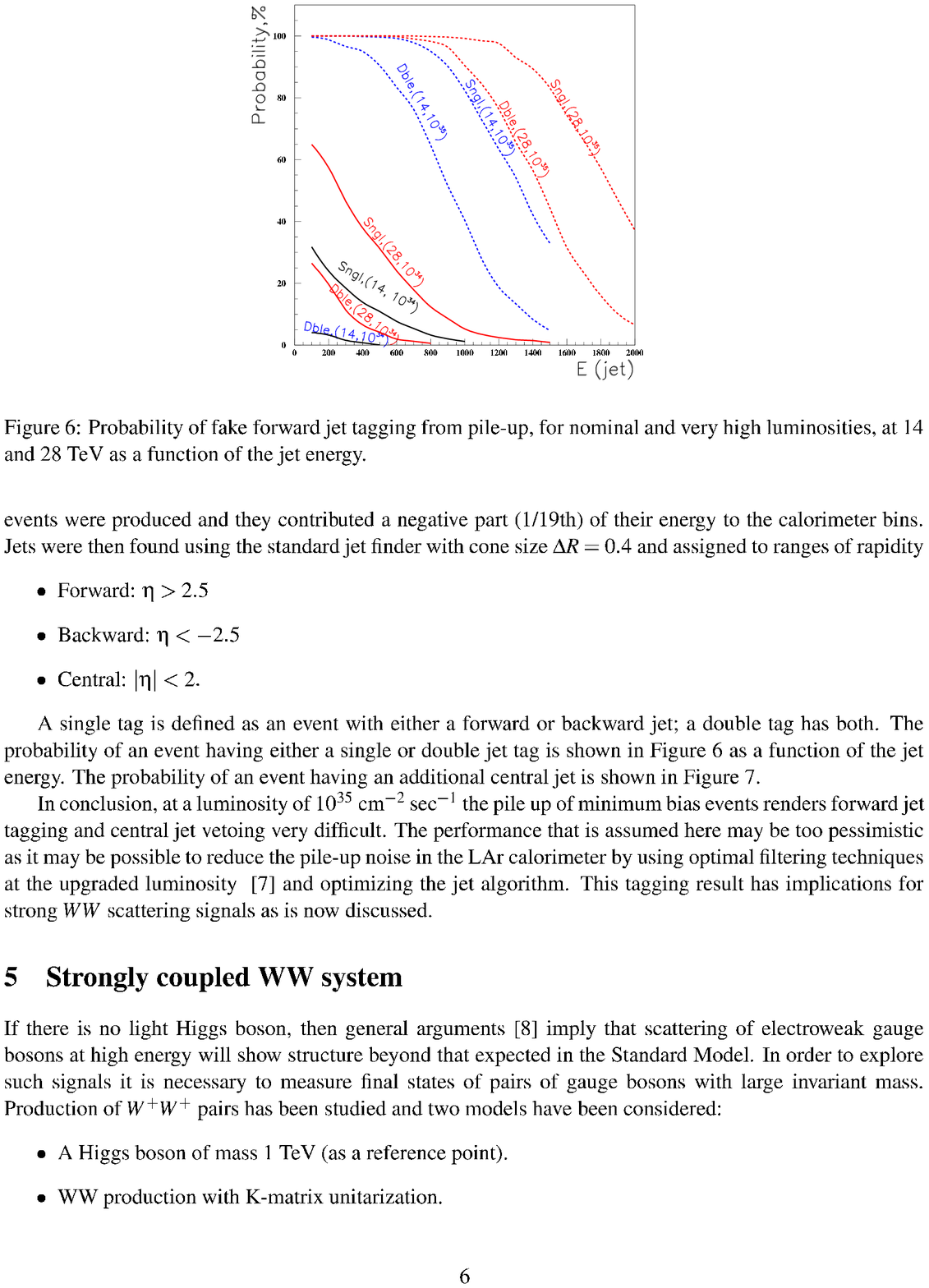}
%    \vspace{-0.2cm}
    \caption{Probability of single and double fake forward jet tagging
from pile-up at ${\cal L}=10^{34}\,{\rm cm}^{-2}\,{\rm s}^{-1}$ (solid) 
and ${\cal
L}=10^{35}\,{\rm cm}^{-2}\,{\rm s}^{-1}$ (dashed) for $pp$ collisions at
$\sqrt{s}=14$~TeV and $\sqrt{s}=28$~TeV as a function of the jet energy
(in GeV). The figure is taken from Ref.~\cite{atlasupgr}.}
    \label{fig:eight}
%    \vspace{-0.4truecm}
  \end{minipage}
   \hspace{0.2truecm}
   \begin{minipage}[b]{.46\linewidth}
    \hspace{0.5truecm}
\includegraphics[height=3.3in,clip,bb=190 500 430 
720]{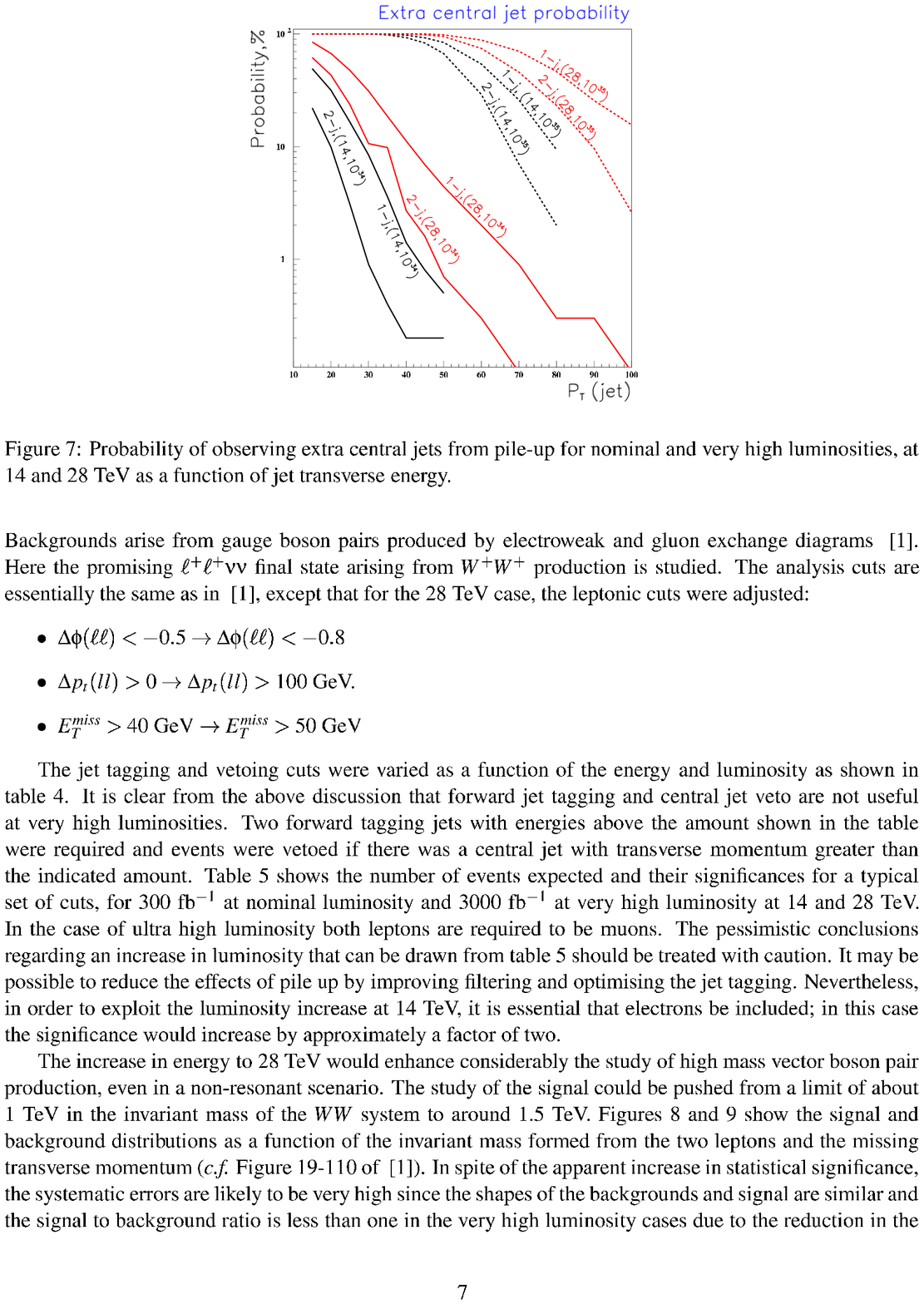}
%    \vspace{0.3cm}
    \caption{Probability of observing one or two extra central jets from
pile-up 
for ${\cal L}=10^{34}\,{\rm cm}^{-2}\,{\rm s}^{-1}$ (solid) and ${\cal
L}=10^{35}\,{\rm cm}^{-2}\,{\rm s}^{-1}$ (dashed) at $\sqrt{s}=14$~TeV and
$\sqrt{s}=28$~TeV as a function of jet transverse energy (in GeV). The
figure is taken from Ref.~\cite{atlasupgr}. }
    \label{fig:nine}
%    \vspace{0.35truecm}
  \end{minipage}
\end{figure}

An integrated luminosity of several ab$^{-1}$ can realistically only be
achieved if the VLHC can be operated at a luminosity close to ${\cal
L}=10^{35}\,{\rm cm}^{-2}\,{\rm s}^{-1}$. Jet tagging and central jet
veto requirements, which are essential for WBF processes such as $qq\to
qqHH$, are expected to become less useful with increasing
luminosity as the pile-up of additional events can cause jets to appear
in these regions and degrade the jet measurements. This was studied in
detail in Ref.~\cite{atlasupgr} for the SLHC. The results are summarized
in Figs.~\ref{fig:eight} and~\ref{fig:nine}.
For the LHC design luminosity of ${\cal L}=10^{34}\,{\rm cm}^{-2}\,{\rm
s}^{-1}$, the background from fake forward jets is small for
$E(j)>300$~GeV and a jet veto cut of $p_T(j)<30$~GeV is feasible. 
For ${\cal L}=10^{35}\,{\rm cm}^{-2}\,{\rm s}^{-1}$, 
the minimum jet energy has to be increased beyond $E(j)\approx 1$~TeV,
and additional strategies would probably have to be employed in order
to keep the background from fake forward jets manageable. In addition,
the jet veto transverse momentum threshold has to be raised to
$70-100$~GeV. For smaller values, the probability for fake central jets
from event pile-up is essentially 100\%. Combined, the increased minimum
energy of the tagging jets, and the higher jet veto $p_T$ threshold
compensate the increased number of events at the higher
luminosity~\cite{atlasupgr}. Similar conclusions are expected to hold
for VLHC energies.

\subsection{Supersymmetry}

\subsubsection{SUGRA Models}

If supersymmetry is connected to the hierarchy problem, it is expected
that sparticles will be sufficiently light that at least some of them
will be observable at the LHC. As the sparticle masses rise, the fine
tuning problem of the SM reappears. If supersymmetry is also the
solution to the dark matter problem, the stable lightest supersymmetric
particle (LSP) is the particle that pervades the universe. This
constraint can be applied to the minimal SUGRA model in order to constrain the
masses of the other sparticles. Recently, minimal SUGRA points have been
proposed~\cite{sugra} which satisfy existing constraints, including the
dark matter constraint, but do not impose any fine tuning limits. Most
of the allowed parameter space corresponds to the case where the
sparticle masses are less than about 1~TeV and thus is accessible to the
LHC. For an integrated luminosity of 100~fb$^{-1}$, squarks and gluinos
with masses up to about 2~TeV can be discovered at the LHC. For
1~ab$^{-1}$, the mass reach improves by $15-30\%$, while doubling the
energy of the LHC would also double the mass reach for squarks and
gluinos to $\approx 4$~TeV~\cite{atlasupgr}. These limits are
essentially model independent.

However, there are regions of parameter space where the annihilation
rate for the LSP can be increased; in these regions the sparticle masses
can be much larger. Two of these points in parameter space, K and L, are
discussed in Ref.~\cite{ian}. Here, we concentrate on point~M which is
characterized by very large sparticle masses: except for the $h^0$ and
$\chi_1^0$, all sparticles have masses larger than
1.2~TeV~\cite{sugra,ian}. We consider a luminosity upgraded LHC and the
VLHC-I with $\sqrt{s}=40$~TeV. For the purposes of this simulation, we assume
that the detector performance at ${\cal L}=10^{35}\,{\rm cm}^{-2}\,{\rm
s}^{-1}$ is the same as that of the ATLAS detector at the LHC design
luminosity. Additional pile-up is taken into account by raising some of
the cuts. Only the $\bar tt$, $Wj$ and $Zj$ backgrounds are included in
the study reported here. Events are selected with hadronic jets and
missing $E_T$ and the following scalar quantity formed
\begin{equation}
M_{eff}=E\llap/_T+\sum_{jets} E_T(jet)+\sum_{leptons} E_T(lepton),
\end{equation}
where the sum extends over all jets with $E_T>50$~GeV and $|\eta|<5$, and
isolated leptons with $E_T>15$~GeV and $|\eta|<2.5$. At least two jets
with 
\begin{equation}
p_T(j)>0.1\,M_{eff}, \quad E\llap/_T>0.3\, M_{eff},\quad
\Delta\phi(j_1,E\llap/_T)<\pi-0.2, \quad {\rm and} \quad
\Delta\phi(j_1,j_2)<{2\pi\over 3}
\end{equation}
are required. 

\begin{figure}
\begin{tabular}{cc}
\includegraphics[width=3.25in]{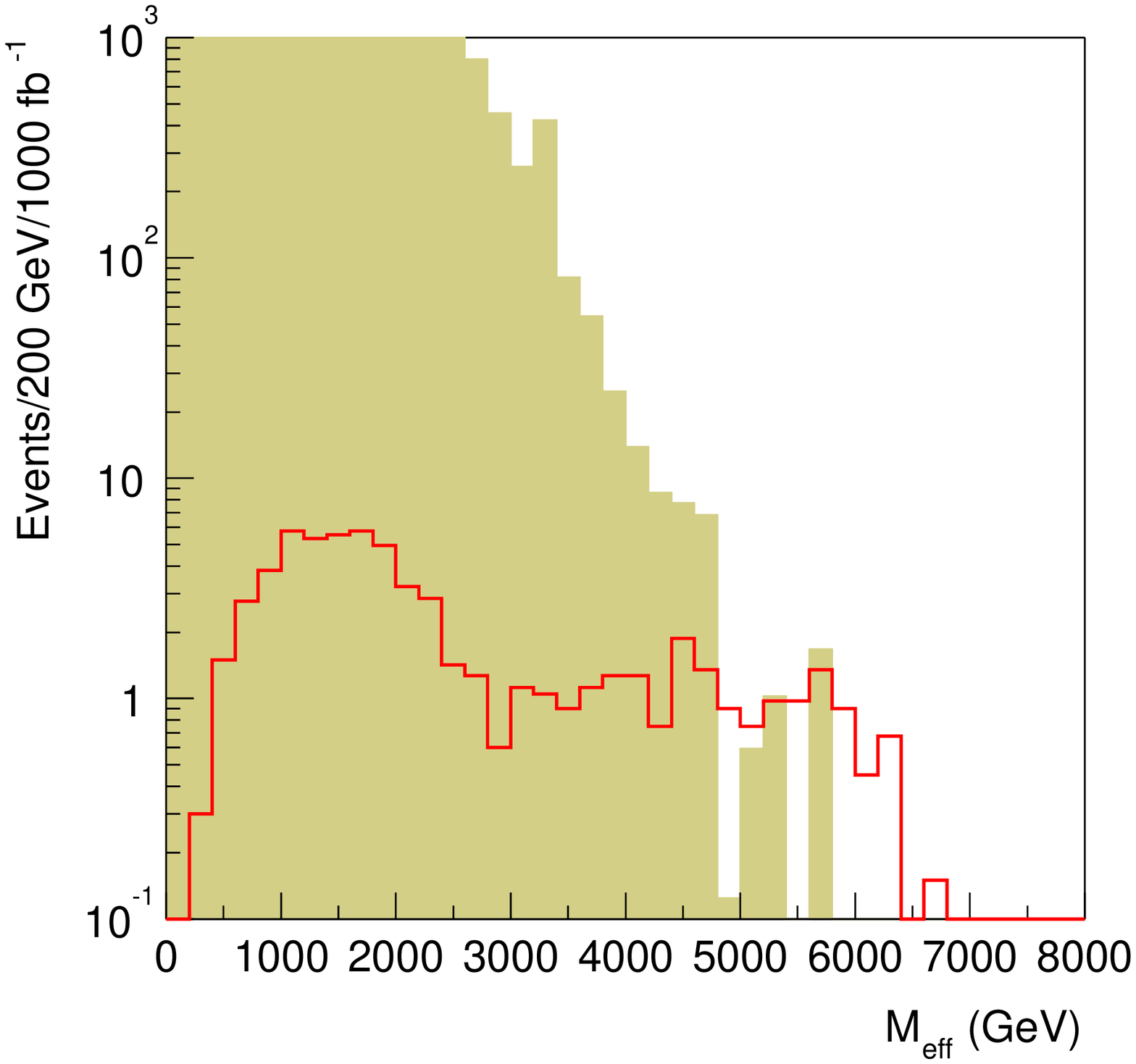} &
\includegraphics[width=3.25in]{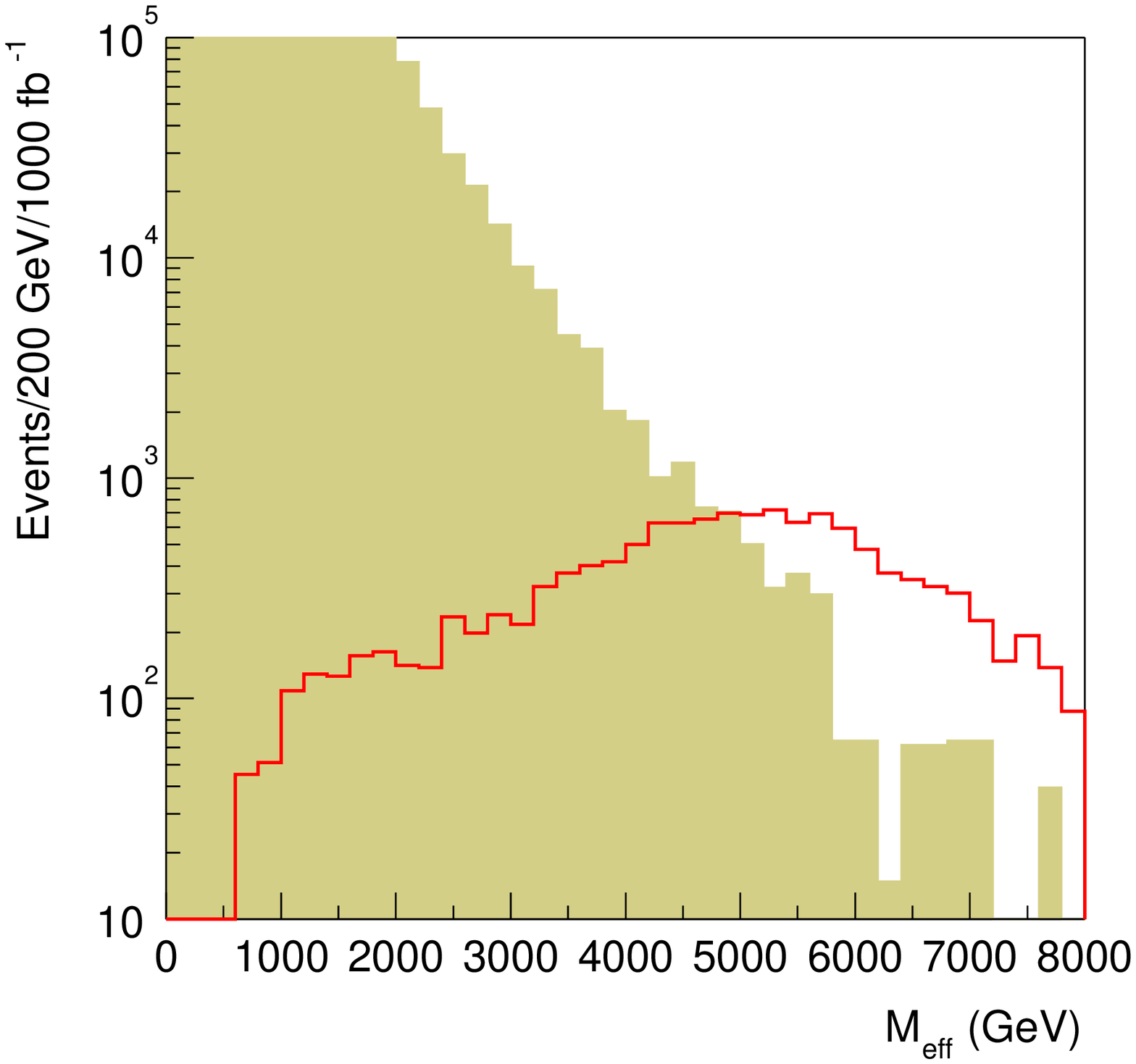} 
\end{tabular}
\caption{$M_{eff}$ distribution for SUGRA point~M at the SLHC (left) and
the VLHC-I (right). The solid (shaded) histogram represents the signal
(SM background).}
\label{fig:ten}
\end{figure}

Due to the large sparticle masses for point~M, 
only 375 supersymmetric events are produced at the LHC even with an 
integrated luminosity of 1~ab$^{-1}$. At the VLHC-I, the cross section
is about a factor~200 larger. The $M_{eff}$ distributions at the SLHC
and VLHC-I are shown in Fig.~\ref{fig:ten}. 
The SLHC signal is very marginal. For $M_{eff}>5$~TeV, there only 6.4
signal events with 3.5 background events. The VLHC-I signal is clearly
visible and could be further optimized. 

\subsubsection{Inverted Mass Hierarchy Models}

\begin{figure}
\includegraphics[width=3.5in]{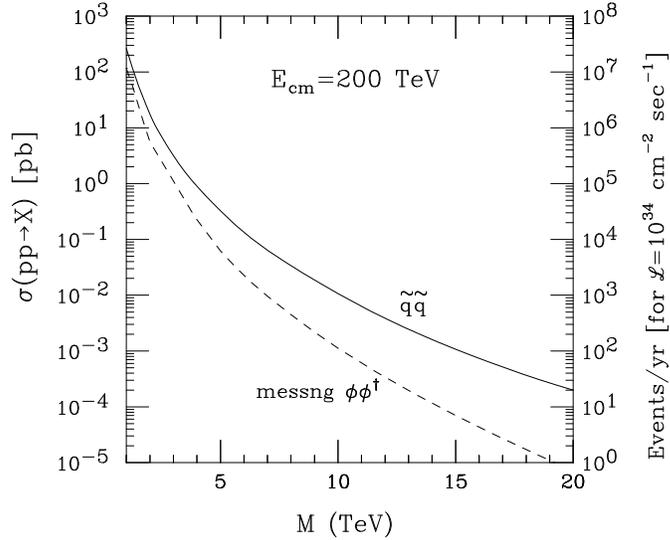}%
\caption{The cross section for squark pair production (solid line) and
pair production of vector-like messenger fields (dashed line) in $pp$
collisions for $\sqrt{s}=200$~TeV. The calculation assumes four
degenerate squark flavors, and two degenerate flavors for the $\Phi$
fields. }
\label{fig:eleven}
\end{figure}

Taking arbitrary weak scale soft supersymmetry breaking parameter
choices generally leads to conflict with various low energy constraints
associated with flavor changing neutral currents (FCNC). Three possibilities
have emerged for building models consistent with low energy constraints:
\begin{enumerate}
\item {\sl universality} of scalar lepton masses. The MSUGRA model
adopts universality as an {\it ad hoc} assumption.
\item {\sl alignment} of fermion and sfermion mass matrices, and
\item {\sl decoupling}, which involves setting sparticle masses to such
high values that loop effects from sparticles are suppressed relative to
SM loops~\cite{dine}.
\end{enumerate}
It is important to notice that ``naturalness'' arguments, which
generally require sparticle masses $<1$~TeV most directly apply to third
generation superpartners, owing to their large Yukawa couplings. In
contrast, the FCNC constraints mainly apply to the scalar masses of the
first two generations. This observation has motivated the construction
of so-called inverted mass hierarchy (IMH) models, where the squarks and
sleptons associated with the first two generations have multi-TeV
masses, while third generation scalars have sub-TeV masses. Models where
the IMH is already in place at the grand unification scale~\cite{baer}
typically have first and second generation scalar masses in the
$5-20$~TeV range and rather heavy charginos, neutralinos and gluinos,
making their discovery a considerable challenge
 at the LHC and a LC. The cross
section for squark pair production (assuming four degenerate squark
flavors) for $\sqrt{s}=200$~TeV as a function of the squark mass is
shown in Fig.~\ref{fig:eleven} (solid line). Assuming an integrated 
luminosity of 200~fb$^{-1}$, it should be possible to discover squarks
in IMH models with masses up to about 20~TeV. 

\subsubsection{Probing the Dynamics of Supersymmetry Breaking at the
VLHC} 
\label{sec:messenger}

Any supersymmetric theory of physics beyond the SM must contain a
mechanism for breaking supersymmetry and a method (messenger) for
communicating this breaking to the superpartners of the SM
particles. Hence, any discovery of supersymmetry immediately implies the
existence of two, possibly distinct, new scales: the fundamental scale
of supersymmetry breaking given by the dimension--two
vacuum expectation value $F$, and the messenger scale $M$. A typical
superpartner mass $\widetilde m$ is related to $M$ and $F$ by~\cite{anderson}
\begin{equation}
\widetilde m\sim\eta\,{F\over M},
\label{eq:susy}
\end{equation}
where $\eta$ is a possible suppression factor originating from
dimensionless couplings. In SUGRA models, $M=M_{Pl}\approx 10^{19}$~GeV and
$\sqrt{F}\sim 10^{10}$~GeV. 

If supersymmetry breaking is communicated by gauge interactions (GMSB
models), $M$ is
replaced by the mass of heavy messenger fields, $\Phi$, and
the parameter $\eta$ contains a factor $\alpha/4\pi$. The messenger
fields form a vector-like representation of the gauge group. If 
$\sqrt{F}\sim M$ and $\widetilde m\stackrel{<}{\sim} 1$~TeV, it is possible 
that the mass of the $\Phi$ fields is
in the range of $10-100$~TeV. It may thus be possible to directly
produce the messenger fields at a VLHC. The cross section for 
pair production of colored spin-0 messenger fields at $\sqrt{s}=200$~TeV
is given by the dashed line in 
Fig.~\ref{fig:eleven}. For the VLHC-II design luminosity of ${\cal
L}=2\cdot 10^{34}\,{\rm cm}^{-2}\,{\rm s}^{-1}$ at this
energy~\cite{vlhcdesign}, it is clear that it will be difficult to
discover messenger fields which are heavier than about 12~TeV. An
additional factor of $\sim 5$ in luminosity would increase the maximum
$\Phi$ mass which can be accessed to $\sim 18$~TeV.

If the messenger fields do not directly couple to the SM fields, the
``messenger number'' is conserved, and the lightest messenger particle
(LMP), $\Phi_0$, is stable. In this scenario, the LMP has
to be lighter than a few~TeV in order not to overclose the universe
in the standard inflationary cosmology~\cite{tao2}. The heavier
messenger fields 
decay into the LMP and a SM gauge field, $V=W,\,Z,\,\gamma,\,g$, 
\begin{equation}
\Phi\to\Phi_0+V\to E\llap/_T+V.
\end{equation}
$\Phi$ pair production thus is signaled by pair production of SM gauge
fields accompanied by a large amount of missing transverse energy from
the LMP's escaping the detector. 

If the messenger fields couple to SM multiplets~\cite{tao}, the LMP can decay,
e.g. via 
\begin{equation}
\Phi_0\to e\mu,~\tilde q_i\tilde q_j,~h\tilde q.
\end{equation}
$\Phi_0\bar\Phi_0$ production thus can result in $e^+e^-\mu^+\mu^-$
events with a bump in the invariant mass of the $e\mu$ systems. The
decays into two squarks, or a Higgs boson and a squark, together with the
subsequent $h$ or $\tilde q$ cascade decays, can lead to
events with very high particle multiplicities in the final state. 

The scales $\sqrt{F}$ and $M$ can be determined experimentally. 
In GMSB models, the gravitino, $\tilde G$, is the lightest supersymmetric
particle and its mass is very small. The
phenomenology of the superpartners of the SM gauge and fermion fields is
then determined by the nature of the next lightest supersymmetric
particle (NLSP), and by its lifetime and decay into a $\tilde
G$~\cite{highpt}. The lifetime of the NLSP,
\begin{equation}
c\tau_{NLSP}=100\mu m\left[{100~{\rm GeV}\over
m_{NLSP}}\right]^5\left[{\sqrt{F}\over 100~{\rm TeV}}\right]^4~,
\end{equation}
measures the overall supersymmetry breaking scale and thus is a crucial
parameter in these models. From the sparticle mass spectrum it is then
possible (see Eq.~(\ref{eq:susy})) to determine the messenger scale
$M$. At the LHC, it should be possible to determine $\sqrt{F}$ with a
precision of about 10\%, and $M$ with a precision of $\sim
30\%$~\cite{fabiola}. If experiments at the LHC discover supersymmetry 
and determine that $M<20$~TeV, the messenger fields $\Phi$ could be
directly produced at the VLHC. Such a scenario would provide a
compelling physics case for a hadron collider operating in the 100~TeV
range.

\subsection{Strong Electroweak Symmetry Breaking}

While the existing precision electroweak measurements prefer a light
Higgs boson~\cite{charlton,p1prec}, the possibility of electroweak
symmetry breaking by new strong dynamics at the TeV scale cannot be
excluded. 

\subsubsection{Strongly Interacting Weak Bosons}

The couplings of longitudinally polarized gauge bosons to each other are
fixed at low energy by the nature of the spontaneously broken
electroweak symmetry and are independent of the details of the mechanism
which breaks the $SU(2)_L\times U(1)_Y$ symmetry. Weak boson scattering 
amplitudes calculated from these couplings violate $S$-matrix unitarity
at center of mass energies of $\sim 1.5$~TeV. To cure this problem, new
physics must enter. In the SM and supersymmetric theories, the cure
arises from one or several weakly coupled Higgs bosons. If no Higgs-like
particles exist, new strong dynamics must enter in the weak boson
scattering amplitudes at high energies. 

Various models exist which can be used as benchmarks for new strong
interactions. The basic signal for all these models is an excess of
events over that predicted by the SM for gauge boson pairs at large
invariant masses produced in vector boson scattering processes. In some
models such as technicolor~\cite{techni,liz} or 
topcolor-assisted technicolor~\cite{topcolor}, resonances appear, in
others the new strong dynamics is signaled by a smooth enhancement over
the SM cross section. This more difficult case has been studied in
detail by the ATLAS 
collaboration for $W^\pm W^\pm$ production via WBF at an upgraded
LHC~\cite{atlasupgr} for the K-matrix unitarization
model~\cite{bagger}. In this model, $S$-matrix unitarity is achieved by
replacing the partial wave amplitudes, $a_\ell^I$, by
\begin{equation}
t_\ell^I={a_\ell^I\over 1-ia_\ell^I}~,
\end{equation}
where $I$ is the isospin, and $\ell$ is the angular momentum quantum
number. 

\begin{figure}
   \begin{minipage}[b]{.46\linewidth}
    \hspace{-0.7truecm}
    \includegraphics[height=3.2in,clip,bb=150 400 450 670]{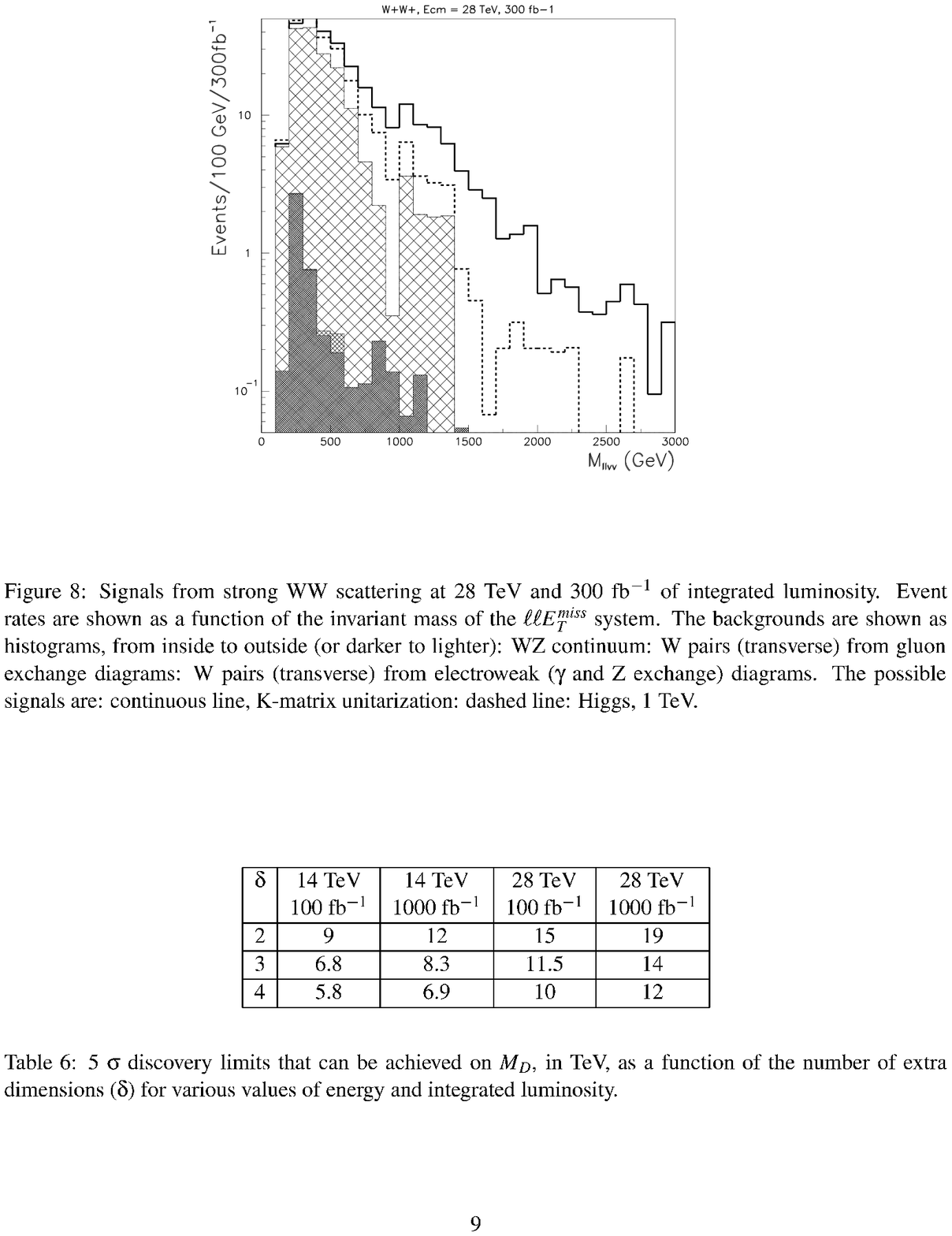}
%    \vspace{-0.2cm}
    \caption{Signals from strong $W^\pm W^\pm$ scattering at
$\sqrt{s}=28$~TeV for an integrated luminosity of 300~fb$^{-1}$. Event
rates are shown as a function of the invariant mass of the $\ell\ell
E\llap/_T$ system. The backgrounds are shown as histograms, from inside
to outside (or darker to lighter): $WZ$ continuum, $W$ pair production
from gluon exchange, and $W$ pair production from photon and $Z$
exchange. The solid line shows the signal from K-matrix
unitarization. For comparison, the dashed line displays the signal for a SM
Higgs boson with $M_H=1$~TeV (from Ref.~\cite{atlasupgr}).}
    \label{fig:twelve}
%    \vspace{-0.4truecm}
  \end{minipage}
   \hspace{0.2truecm}
   \begin{minipage}[b]{.46\linewidth}
    \hspace{0.5truecm}
    \includegraphics[height=3.in]{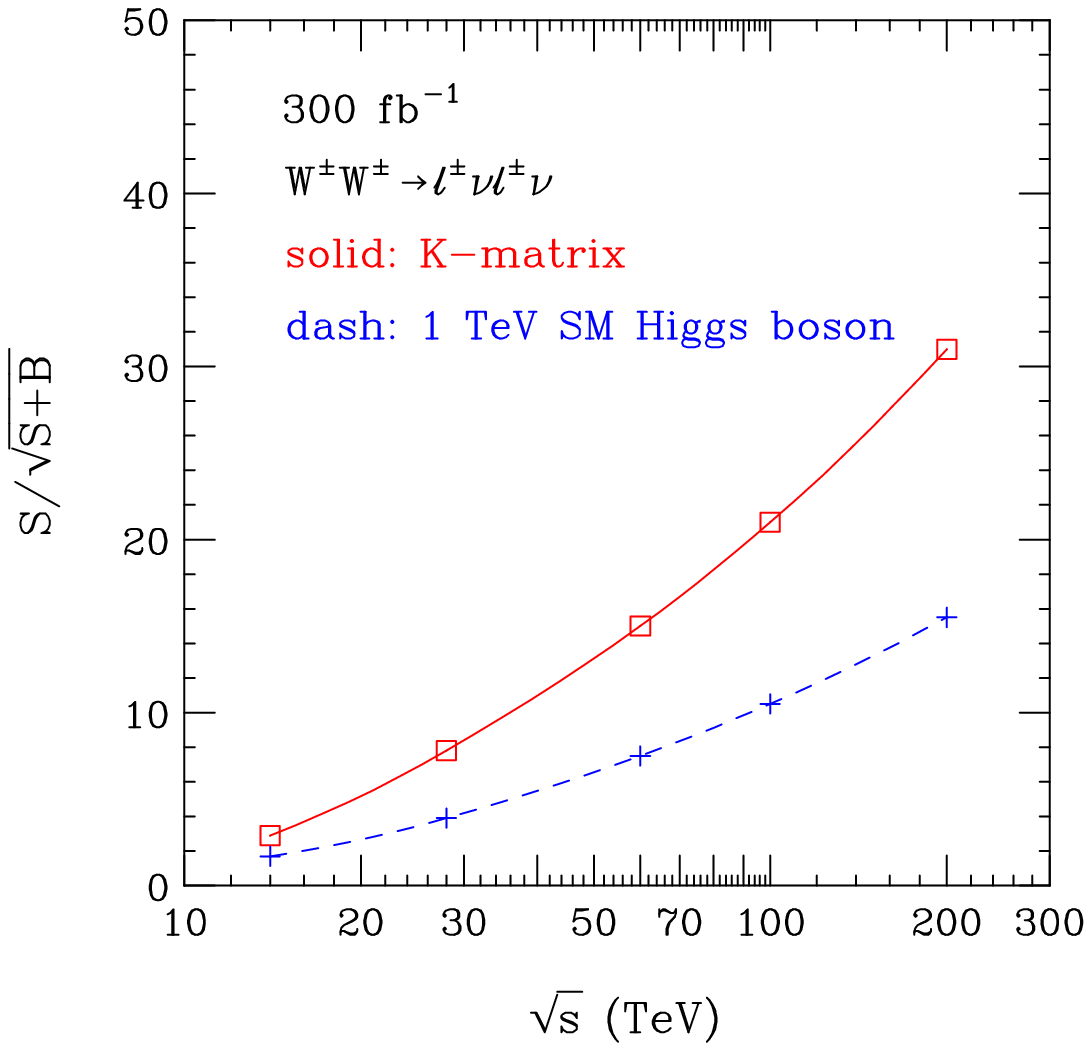}
%    \vspace{0.3cm}
    \caption{Significance of the signal for the K-matrix unitarization
model (solid line) in $W^\pm W^\pm\to\ell^\pm\nu\ell^\pm\nu$ as a
function of the center of mass energy, $\sqrt{s}$, for an integrated
luminosity of 300~fb$^{-1}$. For comparison the result for a SM Higgs
boson with $M_H=1$~TeV (dashed line) in this channel is also shown. }
    \label{fig:thirteen}
    \vspace{1.85truecm}
  \end{minipage}
\end{figure}

Events were selected that have $E\llap/_T>40$~GeV and two same sign
leptons with 
$p_T(\ell)>40$~GeV ($p_T(\ell)>50$~GeV for $\sqrt{s}=28$~TeV) and
$|\eta(\ell)|<1.75$. Backgrounds from $WZ$ and 
$ZZ$ production were rejected by requiring that no third lepton be
present which, in combination with one of the other leptons, is
consistent with the decay of a $Z$ ($|M_Z-m(\ell^+\ell^-)|<15$~GeV). In
addition, the two leptons are required to have $m(\ell\ell)>100$~GeV,
$\Delta\phi(\ell\ell)<-0.5$ ($\Delta\phi(\ell\ell)<-0.8$ for
$\sqrt{s}=28$~TeV), and, for $\sqrt{s}=28$~TeV, $\Delta
p_T(\ell\ell)>100$~GeV. A jet veto requiring no jets with $p_T(j)>50$~GeV
and $|\eta(j)|<2$ is effective against $Wt\bar t$ production. Requiring
two forward jets with $|\eta(j)|>2$ reduces the background from di-boson
production. Jet tagging (vetoing) in the forward (central) region is
essential to extract the signal. 
The resulting signal and background distributions as a
function of the invariant mass formed from the two leptons and the
missing transverse momentum for $\sqrt{s}=28$~TeV are shown in
Fig.~\ref{fig:twelve}. 
Both signal and background distributions are seen to have similar
shapes, which will lead to rather large systematic uncertainties. 

The resulting statistical significances for 300~fb$^{-1}$ as a function
of $\sqrt{s}$ are 
shown in Fig.~\ref{fig:thirteen}. At the LHC, the statistical
significance of the K-matrix unitarization model is marginal
($S/\sqrt{S+B}=2.9$). Increasing the luminosity
of the LHC by a factor~10 above the design luminosity does not improve 
$S/\sqrt{S+B}$. As discussed in
Sec.~\ref{sec:tagging}, the jet veto and jet tagging thresholds have to
be increased substantially in this case, due to the pile-up from the
increased number of interactions per crossing. This effectively
reduces the number of useful events to approximately that present at 
the LHC design luminosity~\cite{atlasupgr}. At the VLHC-I, on the other
hand, the signal stands out clearly ($S/\sqrt{S+B}\approx 11$).  

\subsubsection{Technicolor and Topcolor}

Many models of strong electroweak symmetry breaking predict the
existence of new heavy fermions. Here we briefly discuss two
examples. 

\begin{figure}
\includegraphics[width=3.5in]{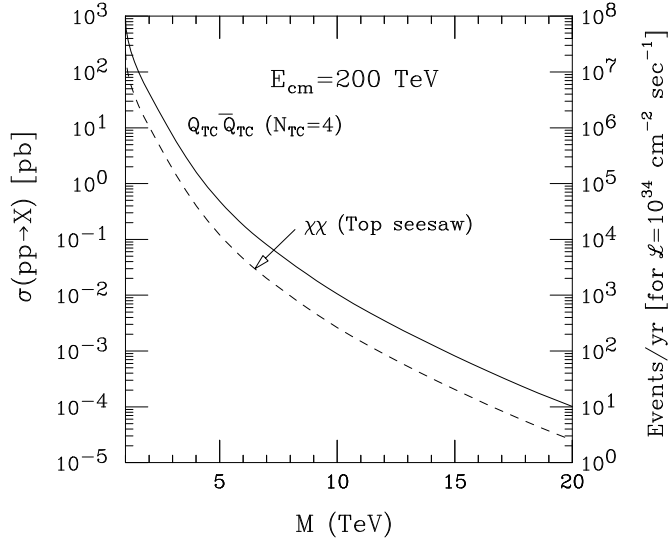}%
\caption{The cross section for $Q_{TC}$ pair production (solid line) and
pair production of $SU(2)_L$ singlet top quark partners $\chi_L$ and
$\chi_R$ in topcolor models (dashed line) in $pp$
collisions at $\sqrt{s}=200$~TeV. The calculation assumes one degenerate
isodoublet of techniquarks, and $\chi_L$ and
$\chi_R$ states which are degenerate in mass. QCD corrections are
simulated by taking into account a $k$-factor of $k=2$. The right
vertical scale shows the number of events per year expected for a
luminosity of ${\cal L}=10^{34}~{\rm cm}^{-2}{\rm s}^{-1}$. }
\label{fig:fourteen}
\end{figure}

In technicolor models~\cite{techni,liz}, the existence of new quarks
bound by novel strong interactions is predicted. A condensate of these 
so-called techniquarks, $Q_{TC}$, breaks the $SU(2)_L\times U(1)_Y$
symmetry. Techniquarks can be pair produced in hadronic collisions via
ordinary strong interactions. Assuming that the techniquarks are in the
fundamental representation of the group $SU(4)_{TC}$ which is supposed to
dynamically break the electroweak symmetry, we show the cross section
for $Q_{TC}$ pair production as a function of the $Q_{TC}$ mass for
$\sqrt{s}=200$~TeV in Fig.~\ref{fig:fourteen} (solid line). 
It should be possible to detect techniquarks with masses up to about
15~TeV at a VLHC-II with an integrated luminosity of 100~fb$^{-1}$. 

The top-seesaw model~\cite{topseesaw1,topseesaw2}, a variant of the
original~\cite{topcolor} topcolor model, predicts the existence of
partners of the top quark, $\chi_R$ and $\chi_L$, which transform as
singlets under $SU(2)_L$. These fermions are responsible for the
dynamical breaking of the electroweak symmetry. 
Current precision data constrain their masses
to the range $3~{\rm TeV}\stackrel{<}{\sim}M_\chi\stackrel{<}{\sim} 
10$~TeV~\cite{topseesaw2}. The cross section for $\chi$ pair production
in $pp$ collisions at $\sqrt{s}=200$~TeV is given by the dashed line in
Fig.~\ref{fig:fourteen}, assuming that the $\chi_L$ and $\chi_R$ are
degenerate in mass. For $M_\chi=10$~TeV and an integrated luminosity of
100~fb$^{-1}$, approximately 100~events are expected. If the electroweak
symmetry is dynamically broken via the mechanism suggested by top-seesaw
models, the VLHC-II will be able to directly produce the particles which
trigger the breaking of $SU(2)_L\times U(1)_Y$. 

\subsection{Extra Gauge Bosons}
\label{sec:extragauge}

The existence of heavy neutral ($Z'$) and/or charged ($W'$) vector
bosons is a feature of many extensions of the SM. They arise in
extended gauge theories including grand unified theories, superstring
theories, left-right symmetric models, and other models such as the BESS
model~\cite{bess} and models of composite gauge bosons (for a review see
Ref.~\cite{godfrey}). The mass reach depends on the couplings of the
$W'$ and $Z'$ bosons to quark and leptons. 

\begin{figure}
\includegraphics[width=4.5in,clip,bb=13 113 599 779]{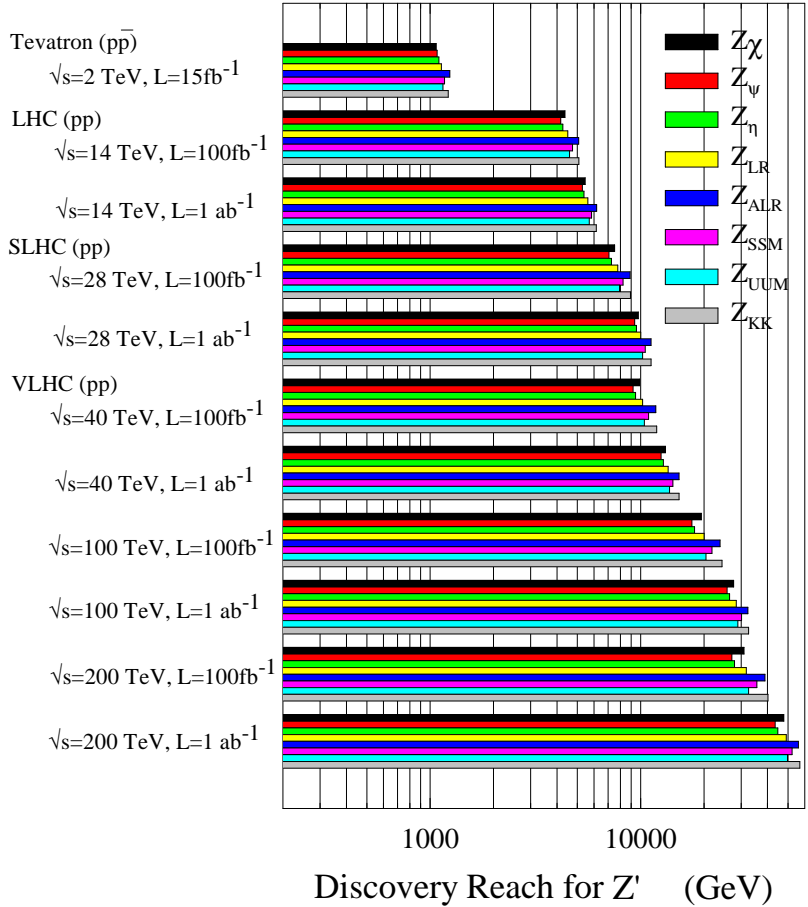}%
\caption{Discovery limits for extra neutral gauge bosons ($Z'$) for a
variety of models. $Z_\chi$, $Z_\psi$ and $Z_\eta$ refer to the $Z'$
bosons predicted by grand unified models of rank~5, such as $SO(10)$ or
$E(6)$~\cite{godfrey}. $Z_{LR}$ is the extra gauge boson of left-right
symmetric models embedded in $SO(10)$~\cite{godfrey}. The alternative
left-right symmetric model of Ref.~\cite{alrm} predicts an extra neutral
gauge boson, $Z_{ALR}$. The $Z_{SSM}$ case corresponds to a $Z'$ with SM
like couplings. The $Z_{UUM}$ is the extra gauge boson of the un-unified
model of Ref.~\cite{uum}. Finally, the $Z_{KK}$ refers to the lowest
Kaluza-Klein excitation of the SM $Z$ boson which appears in models with
TeV-scale extra dimensions (see Sec.~\ref{sec:xd}). 
The discovery limits are based on 10 events in the $e^+e^-$ and
$\mu^+\mu^-$ channels (from Ref.~\cite{god_sn}).}
\label{fig:fifteen}
\end{figure}

The search limits for $Z'$ bosons at the Tevatron, LHC, SLHC and
VLHC for various integrated luminosities are shown in
Fig.~\ref{fig:fifteen}~\cite{god_sn}. These bounds are based on requiring 10
$Z'\to\ell^+\ell^-$ ($\ell=e,\,\mu$) events and are in good agreement
with limits obtained by CDF~\cite{cdfzp} and D\O~\cite{d0zp}.
The discovery limits in general improve by about 20\% if the integrated
luminosity is increased from 100~fb$^{-1}$ to 1~ab$^{-1}$. Note that the
search reaches of a SLHC with $\sqrt{s}=28$~TeV and 1~ab$^{-1}$ and a 
VLHC-I with $\sqrt{s}=40$~TeV and 100~fb$^{-1}$ are similar. For
$\sqrt{s}=100$~TeV ($\sqrt{s}=200$~TeV) and 100~fb$^{-1}$, $Z'$ bosons
with masses up to $20-25$~TeV ($30-40$~TeV) can be
discovered. Qualitatively similar results are obtained for $W'$ bosons. 
At an $e^+e^-$ or muon collider, direct searches for a
$Z'$ boson are limited to the region $M_{Z'}<\sqrt{s}$. 

\subsection{Compositeness}

Composite models of quarks and leptons attempt to overcome the
shortcomings of the SM by assuming that they are bound states of more
fundamental fermions, and perhaps bosons, bound together by a new
interaction which is characterized by an energy scale $\Lambda$. At
energies much smaller than $\Lambda$, the substructure of quarks and
leptons is signalled by the appearance of four fermion contact
interactions which arise from the exchange of bound states of the
subconstituents~\cite{ELP}. For energies similar to or larger than
$\Lambda$, one expects that excited states of the known quarks and
leptons are produced~\cite{qstar1,qstar2}.

\subsubsection{Contact Interactions}

\begin{figure}
\includegraphics[width=3.5in,clip,bb=170 510 420 720]{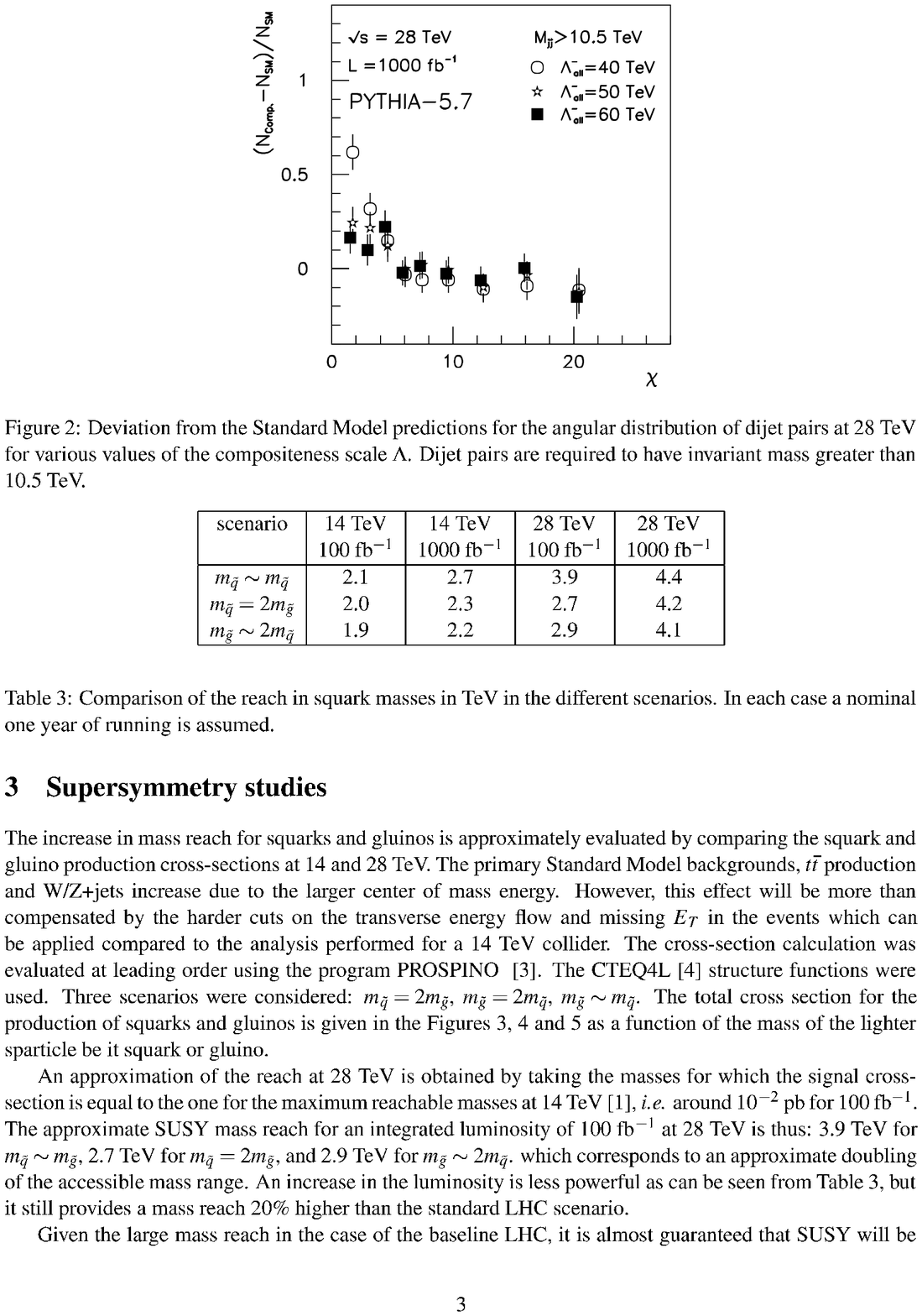}%
\caption{Deviation from the SM prediction for the angular distribution
of dijet pairs at $\sqrt{s}=28$~TeV for various values of
$\Lambda$. Dijet pairs are required to have invariant mass
$M_{jj}>10.5$~TeV (from Ref.~\cite{atlasupgr}). Constructive interference
between SM and contact terms is assumed in the matrix elements. }
\label{fig:sixteen}
\end{figure}

The lowest order contact terms are dimension~6 four-fermion interactions
which can affect jet and Drell-Yan production at hadron
colliders. Compared with the SM terms, they are suppressed by a factor
$1/\Lambda^2$. The signature for four quark contact interactions, for
example, would be an excess of events at large transverse energy,
$E_T$. Another 
signal for quark -- quark contact interactions, which, in contrast to
the $E_T$ distribution, is not sensitive to theoretical or jet energy
uncertainties, is the dijet angular distribution, $d\sigma/d\chi$, where
\begin{equation}
\chi={1+|\cos\theta|\over 1-|\cos\theta|}
\end{equation}
with $\theta$ being the angle between a jet and the beam in the center
of mass of the dijet system. If contact terms are
present, the dijet angular distribution is more isotropic than that
predicted by QCD. Figure~\ref{fig:sixteen} shows the deviation of the angular
distribution from the SM prediction at an upgraded LHC.
The maximum compositeness scales which can be probed using
$d\sigma/d\chi$ are shown in Table~\ref{tab:two}. No detailed study of
what values of $\Lambda$ could be probed at a VLHC has been carried out
so far. Extrapolating the limits listed in Table~\ref{tab:two} to
$\sqrt{s}=200$~TeV, one finds that one should be able to probe values of
$\Lambda\approx 100$~TeV at the VLHC-II with an integrated luminosity
of 300~fb$^{-1}$. 
\begin{table}
\caption{The 95\% confidence level limits that can be obtained for
$\Lambda$ using the dijet angular distribution (from
Ref.~\cite{atlasupgr}).\\}
\label{tab:two}
\begin{tabular}{|c|c|c|c|c|}
\tableline
energy & $\sqrt{s}=14$~TeV & $\sqrt{s}=14$~TeV & $\sqrt{s}=28$~TeV &
$\sqrt{s}=28$~TeV \\
$\int\!{\cal L}dt$ & 300~fb$^{-1}$ & 3000~fb$^{-1}$ & 300~fb$^{-1}$ &
3000~fb$^{-1}$ \\
\tableline
$\Lambda$ (TeV) & 40 & 60 & 60 & 85\\
\tableline
\end{tabular}
\end{table}

\subsubsection{Excited Quarks}

Conclusive evidence for a new layer of substructure would be provided by
the direct observation of excited states of the known quarks and
leptons. In the following we shall concentrate on excited quarks with
spin~$1/2$ and weak isospin~$1/2$. Excited up- and down-quarks should be
almost degenerate in mass if the dynamics which binds the quark constituents
respects isospin. The coupling between excited spin~1/2
quarks, ordinary quarks and gauge bosons is uniquely fixed to be of
magnetic moment type by gauge invariance~\cite{qstar1},
\begin{equation}
{\cal L}_{eff}={1\over
2m^*}\,\bar{q^*}\sigma^{\mu\nu}\left[g_sf_s{\lambda^a\over
2}F^a_{\mu\nu}+gf{\vec\tau\over 2}\vec W_{\mu\nu}+g'f'{Y\over
2}B_{\mu\nu}\right] q_L + h.c.~.
\end{equation}
Here, $q^*$ and $q_L$ denote the isospin doublets of excited and
lefthanded ordinary quarks, $V_{\mu\nu}$, $V=F^a,\,\vec W,\,B$, is the
field strength tensor for the gluon, the $SU(2)$ and the $U(1)$ gauge
fields, and $Y=1/3$ is the weak hypercharge. Finally, $g_s$, $g$ and
$g'$ are the gauge couplings and $f_s$, $f$ and $f'$ are free parameters
determined by the composite dynamics. Naively one would expect that they
are all of ${\cal O}(1)$. To set the scale in ${\cal L}_{eff}$ we choose
the $q^*$-mass $m^*$. 

Excited quarks decay into quarks and a gluon, photon or $W/Z$ boson, or,
via contact interactions into $\bar qqq'$ final
states~\cite{qstar2}. Subsequently, only decays via gauge interactions
are considered. Excited quarks are then expected to decay predominantly
via strong interactions; $q^*\to q\gamma$, $q^*\to q'W$ and $q^*\to qZ$
will typically appear at the few per cent level.

\begin{figure}
   \begin{minipage}[b]{.46\linewidth}
    \hspace{-0.7truecm}
    \includegraphics[height=3.in]{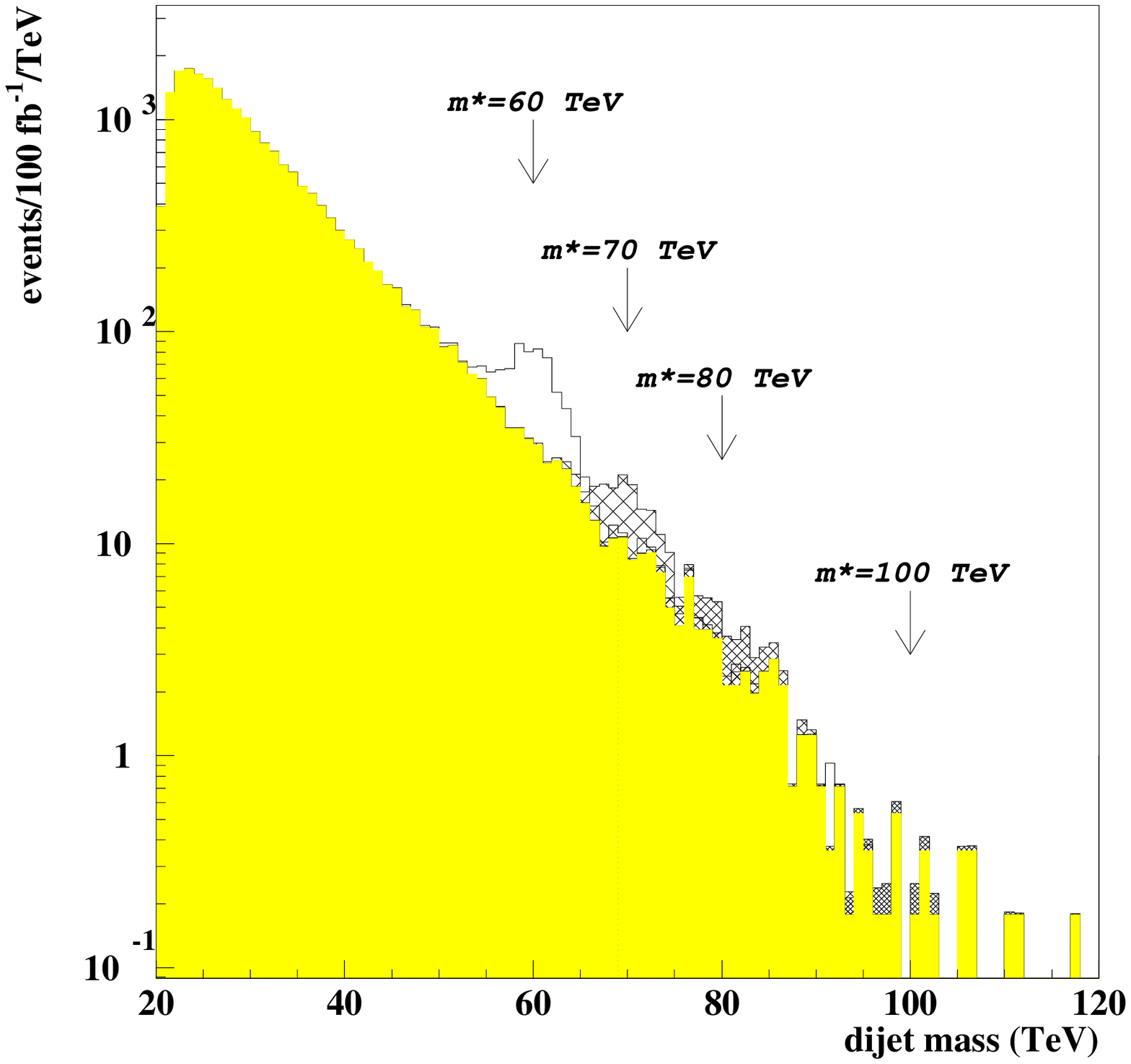}
%    \vspace{-0.2cm}
    \caption{Dijet invariant mass spectrum for $pp$ collisions at
$\sqrt{s}=200$~TeV, assuming an integrated luminosity of
100~fb$^{-1}$. Four excited quark resonances are shown above the QCD
dijet background (shaded). To simulate detector response, a constant term
$C=0.05$ in the jet energy resolution is assumed.}
    \label{fig:seventeen}
%    \vspace{-0.4truecm}
  \end{minipage}
   \hspace{0.2truecm}
   \begin{minipage}[b]{.46\linewidth}
    \hspace{0.5truecm}
    \includegraphics[height=3.4in]{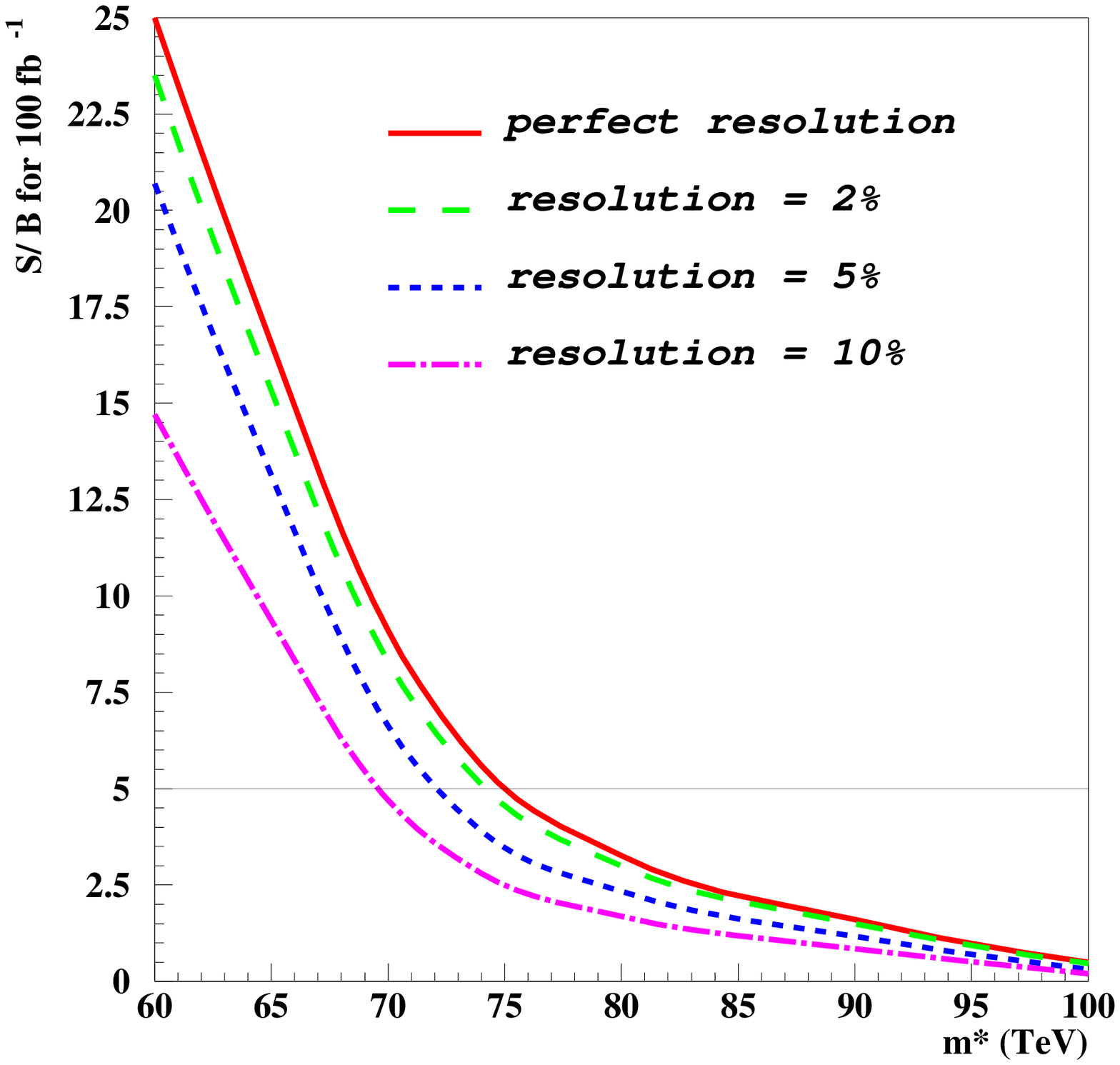}
%    \vspace{0.1cm}
    \caption{Significance $S/\sqrt{B}$ as a function of $m^*$ for $pp$
collisions at $\sqrt{s}=200$~TeV, $\int\!{\cal L}dt=100$~fb$^{-1}$ and four
different values of the jet energy resolution, $C=0$, $C=0.02$, $C=0.05$ and
$C=0.1$.  }
    \label{fig:eighteen}
    \vspace{0.7truecm}
  \end{minipage}
\end{figure}

In hadronic collisions, excited quarks can be produced singly via quark
gluon fusion. The subsequent $q^*\to qg$ decay leads to a peak in the two
jet invariant mass distribution located at $m(jj)=m^*$. At the LHC, with
100~fb$^{-1}$, excited quarks with masses up to $m^*=6.6$~TeV can be
discovered~\cite{atlasupgr,atl_qstar}. For $\sqrt{s}=28$~TeV, the reach
can be increased to $m^*\approx 10$~TeV~\cite{atlasupgr}. At the VLHC, 
much higher masses can be probed. The dijet
invariant mass distribution for $pp$ collisions at $\sqrt{s}=200$~TeV,
assuming $f_s=f=f'=1$ and $m_{u^*}=m_{d^*}$, is shown in 
Fig.~\ref{fig:seventeen}.
{\tt PYTHIA 6.158}~\cite{pythia} has been used for the simulation and 
both jets are required to have $p_T(j)>10$~TeV and
$|\eta(j)|<2$. Detector resolution effects are simulated by taking a
constant term of $C=0.05$ in the jet energy resolution into account. The
sampling term in the jet energy resolution will have a very small effect
at the
energies considered here. The significance, $S/\sqrt{B}$ as a function
of the excited quark mass, $m^*$, is shown in
Fig.~\ref{fig:eighteen}. Assuming a minimum significance of
$S/\sqrt{B}=5$ to claim discovery, one concludes that, depending on the
detector resolution, excited quarks with a mass up to $70-75$~TeV can be
found at a $pp$ collider with $\sqrt{s}=200$~TeV and an integrated
luminosity of 100~fb$^{-1}$. Similar results have been obtained in
Ref.~\cite{harris}. 

The $q^*$ search in the dijet channel is limited mainly by the QCD
background. In the $q^*\to q\gamma$, $q^*\to Wq'$ and $q^*\to Zq$
channels the signal is reduced by the relatively smaller branching
fractions of these channels. However, the backgrounds are reduced as
well. As a result, the $q^*$ mass reach in these channels is similar to
that in the dijet channel. This has been verified by explicit studies
for the LHC~\cite{atl_qW} and SLHC~\cite{atlasupgr}.

Should the (upgraded) LHC discover contact interactions in two jet
production, the VLHC will be able to directly probe the scale of new
physics. 

\subsection{Extra Dimensions}

There is much recent theoretical interest in models of particle physics
that have extra spatial dimensions in addition to the $3+1$ dimensions
of normal space time. In these models, new physics can appear at a mass
scale of ${\cal O}(1$~TeV) and may be accessible at future
hadron colliders. The models considered so far can be grouped in three
classes which lead to very different phenomenologies and collider
signatures. 
\begin{enumerate}
\item The large extra dimension (ADD) scenario~\cite{add} which predicts
the emission and exchange of Kaluza-Klein (KK) towers of gravitons
that are finely spaced in mass. 
\item Models with TeV-scale extra dimensions (TeV)~\cite{anton} which
predict the 
existence of KK excitations of the SM gauge (and possibly other) fields
at the TeV scale.
\item Models with warped extra dimensions such as the model of
Ref.~\cite{rs} (RS) which predict graviton resonances with both weak
scale masses and couplings to matter. 
\end{enumerate}
In the following we briefly discuss these three classes as well as the
production of black holes at hadron colliders. 
%: The signals of models
%with extra spatial dimensions at $e^+e^-$ linear colliders are discussed
%in Ref.~\cite{tgr}.

\subsubsection{Large Extra Dimensions}

\begin{figure}
   \begin{minipage}[b]{.46\linewidth}
    \hspace{-0.7truecm}
    \includegraphics[height=3.in,clip,bb=0 170 530 
700]{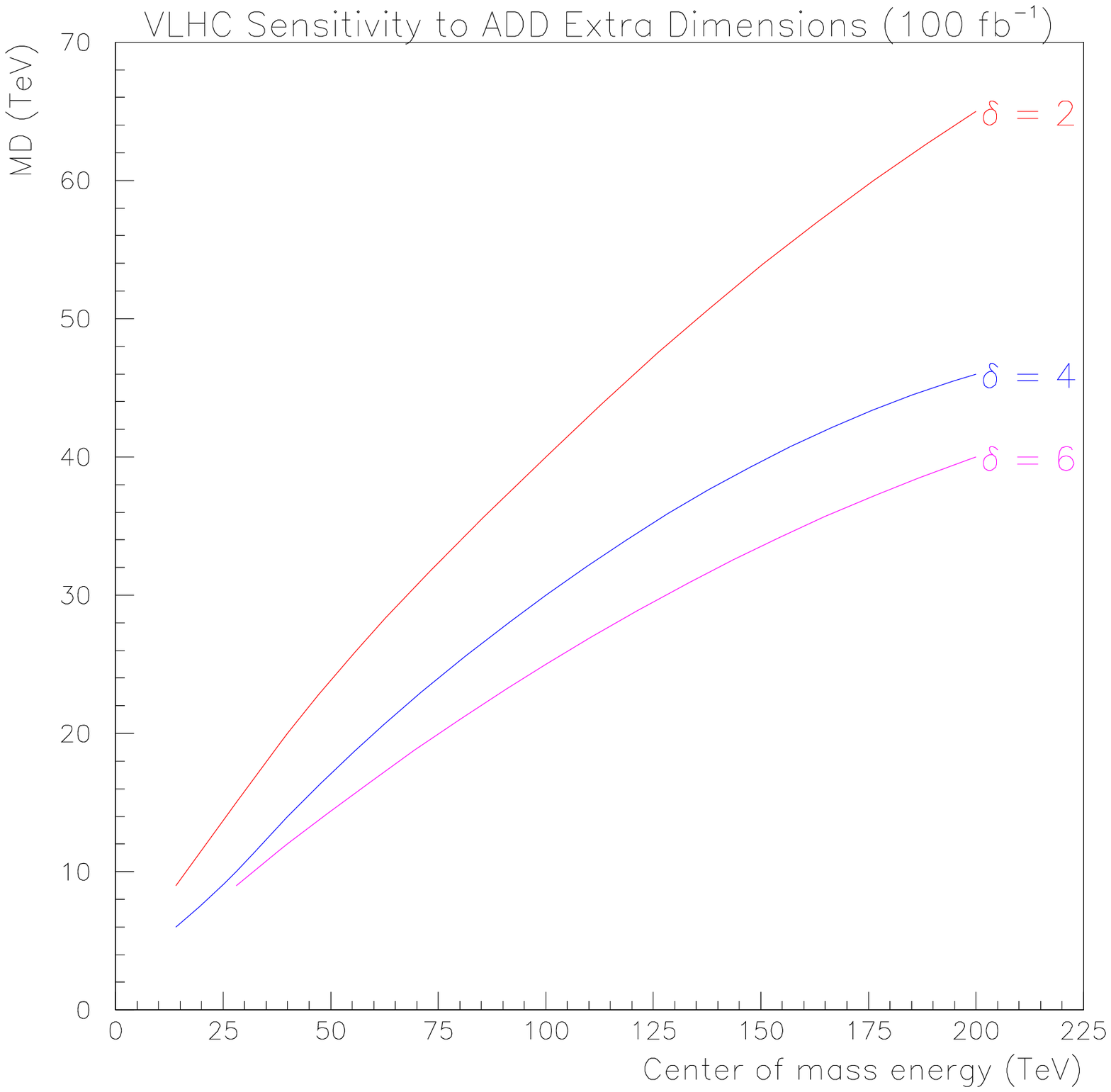}
%    \vspace{-0.2cm}
    \caption{$5\sigma$ discovery limits that can be achieved for $M_D$
for an integrated luminosity of 100~fb$^{-1}$ as a function
of $\sqrt{s}$. Results are shown for 2, 4~and 6~additional dimensions.}
    \label{fig:nineteen}
    \vspace{0.8truecm}
  \end{minipage}
   \hspace{0.2truecm}
   \begin{minipage}[b]{.46\linewidth}
    \hspace{0.5truecm}
    \includegraphics[height=3.2in,angle=90]{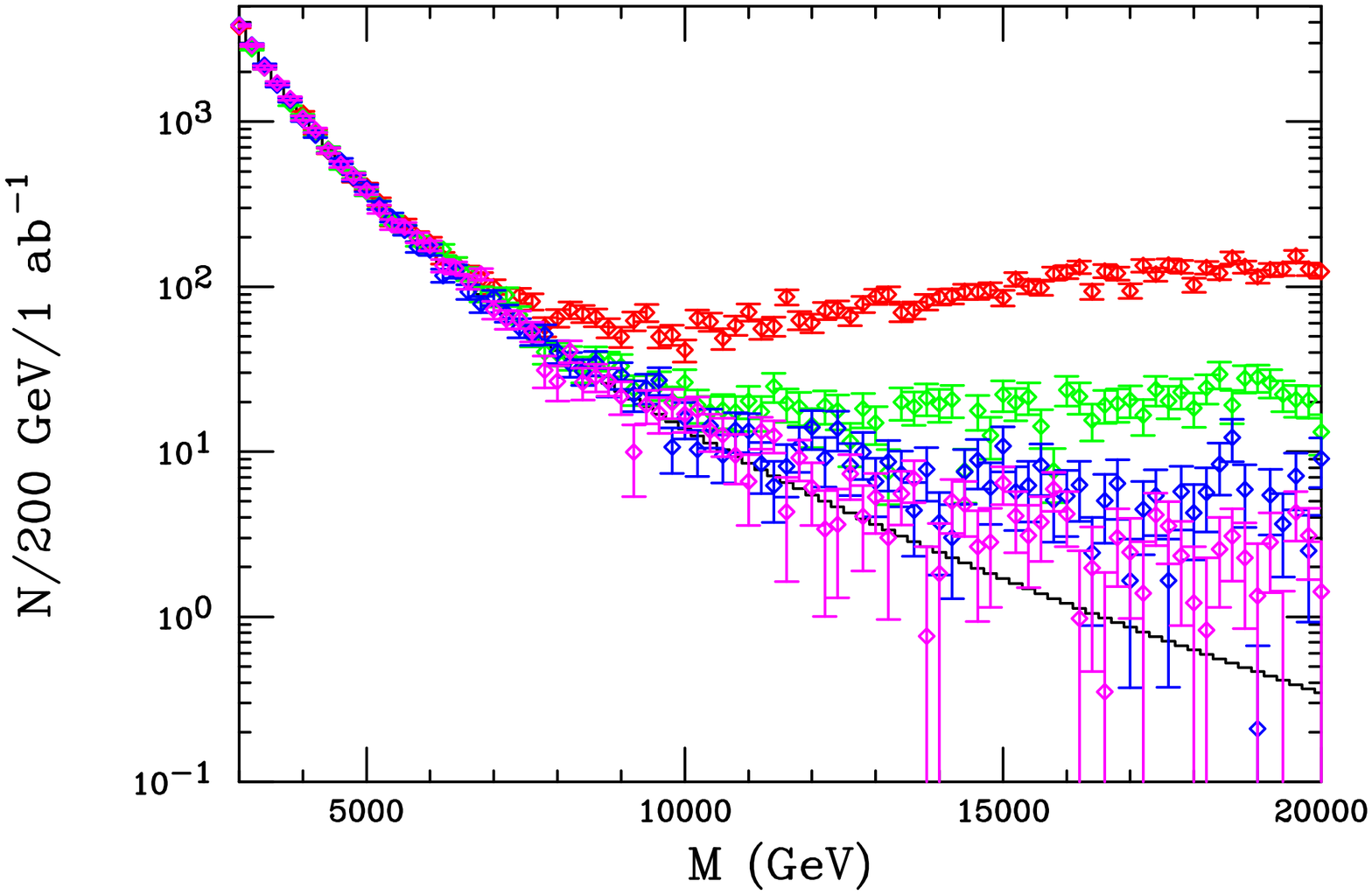}
%    \vspace{0.6cm}
    \caption{Event rate per 200~GeV mass bin for the Drell-Yan process,
$pp\to\ell^+\ell^-$, 
as a function of the di-lepton invariant mass for $\sqrt{s}=200$~TeV and
$\int\!{\cal L}dt=1~{\rm ab}^{-1}$. Both leptons are required to have
rapidity $|\eta(\ell)|<2.5$. The solid histogram is the SM prediction
whereas the 'data' points represent the predictions of the ADD
model. The red, green, blue and magenta points (reading from top to
bottom on the plot) correspond to $M_s=20$,
25, 30 and 35~TeV respectively (from Ref.~\cite{rizzo}). }
    \label{fig:twenty}
  \end{minipage}
\end{figure}

In models with large extra dimensions, massive graviton states can be 
produced in association with jets or photons. Since these graviton states have
gravitational strength couplings, they escape the detector and thus give
rise to missing transverse energy. The properties of the KK graviton
tower are parameterized in terms of the number of additional dimensions,
$\delta$, and the fundamental scale $M_D$. Since the $E\llap/_T+$~jets final
state is more sensitive it will be considered here. The background is
dominated by $Z(\to\bar\nu\nu)+$~jets production. The maximum scale
$M_D$ which can be accessed for $\delta=2$, $\delta=4$ and $\delta=6$ and an 
integrated luminosity of 100~fb$^{-1}$ is shown in 
Fig.~\ref{fig:nineteen} as a function of $\sqrt{s}$.
At $\sqrt{s}=200$~TeV, values of $M_D$ up to 65~TeV (38~TeV) can be
probed for $\delta=2$ ($\delta=6$). This is a factor $7-8$ higher than
the scale which can be probed at the LHC with the same integrated
luminosity~\cite{atlasupgr,vac}. 

One can also search for indirect effects of the KK graviton towers
appearing in ADD models, for example in Drell-Yan
production~\cite{hewett}. In the SM, the Drell-Yan reaction is a result
of photons and $Z$ bosons mediating the process $\bar
qq\to\ell^+\ell^-$ ($\ell=e,\,\mu$). In the ADD model, graviton towers
can also be exchanged, and an additional sub-process,
$gg\to\ell^+\ell^-$, mediated solely by gravitons contributes. The
effect of the graviton towers can be described through a set of
dimension~8 operators in the limit that the parton center of mass energy
is much larger than the cut-off scale, $M_s$, which is of order
$M_D$. Current experimental data from LEP and the Tevatron~\cite{addlim}
require that $M_s\geq 1$~TeV, and values of $M_s$ as large as a few tens
of TeV may be conceivable in this framework. The distortion of
the differential cross section of the Drell-Yan process at large values 
of $m(\ell\ell)$ through these dimension~8 operators can probe
such high mass scales in a manner similar to searches for contact
interactions in composite models. The shape of the invariant mass
distribution will tell us that the underlying physics arises from
dimension~8 operators, while the angular distribution of the leptons
at large di-lepton invariant masses would conform to the shape expected
from the exchange of a spin-2 object, confirming the gravitational
origin of the effect. At the LHC, values of $M_s$ up to
7.9~TeV (7.0~TeV) can be probed in Drell-Yan production for 2 (4) extra
dimensions with 100~fb$^{-1}$~\cite{kaba}. The di-lepton invariant mass 
distribution for a
VLHC operating at $\sqrt{s}=200$~TeV in the SM and for various values of
$M_s$ is shown in Fig.~\ref{fig:twenty}. It is clear that values of
$M_s$ up to about 35~TeV can be probed at a VLHC. Similar mass scales
can be probed at CLIC~\cite{tgr}. 

\subsubsection{TeV-scale Extra Dimensions}
\label{sec:xd}

In the simplest versions of TeV-scale theories with extra dimensions,
only the SM gauge fields are in the bulk whereas fermions remain at the
orbifold fixed points. Higgs fields may lie at the fixed points or
propagate in the bulk. In this simplest case with one extra dimension,
to a good approximation, the masses of the KK excited gauge bosons are
given by $M_n=nM_c$, where $M_c$ is the compactification scale. All KK
tower states have identical couplings to the SM fermions. At the LHC,
with 100~fb$^{-1}$ (3~ab$^{-1}$), it will be possible to discover the
lowest lying KK $\gamma,\, Z$ excitation of this model with mass up to
$M_{KK}=5.2$~TeV ($M_{KK}=6.5$~TeV) (see Sec.~\ref{sec:extragauge}). 
However, there is no hope of observing higher excitations. At a VLHC
with $\sqrt{s}=200$~TeV and 100~fb$^{-1}$, the limit can be pushed up to
$M_{KK}\approx 30$~TeV.

\begin{figure}
   \begin{minipage}[b]{.46\linewidth}
    \hspace{-0.7truecm}
    \includegraphics[width=3.1in]{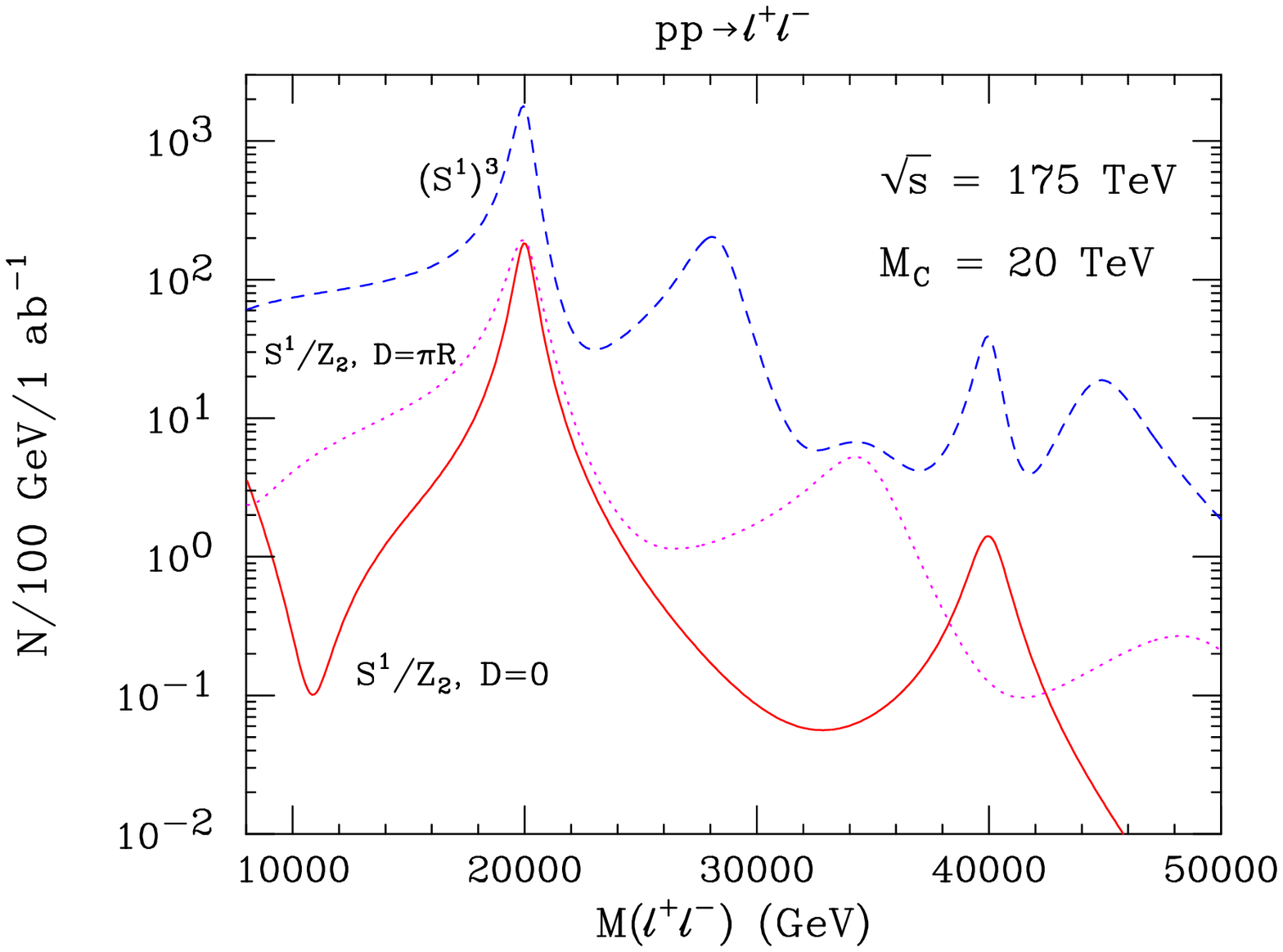}
%    \vspace{-0.2cm}
    \caption{The number of Drell-Yan events expected per 100~GeV bin for an
integrated luminosity of 1~ab$^{-1}$ in several models with two or more
extra dimensions and a compactification scale of $M_c=20$~TeV at the
VLHC. Both leptons are required to have rapidity $|\eta(\ell)|<2.5$. }
    \label{fig:twentyone}
%    \vspace{-0.4truecm}
  \end{minipage}
   \hspace{0.2truecm}
   \begin{minipage}[b]{.46\linewidth}
    \hspace{0.5truecm}
    \includegraphics[width=3.2in]{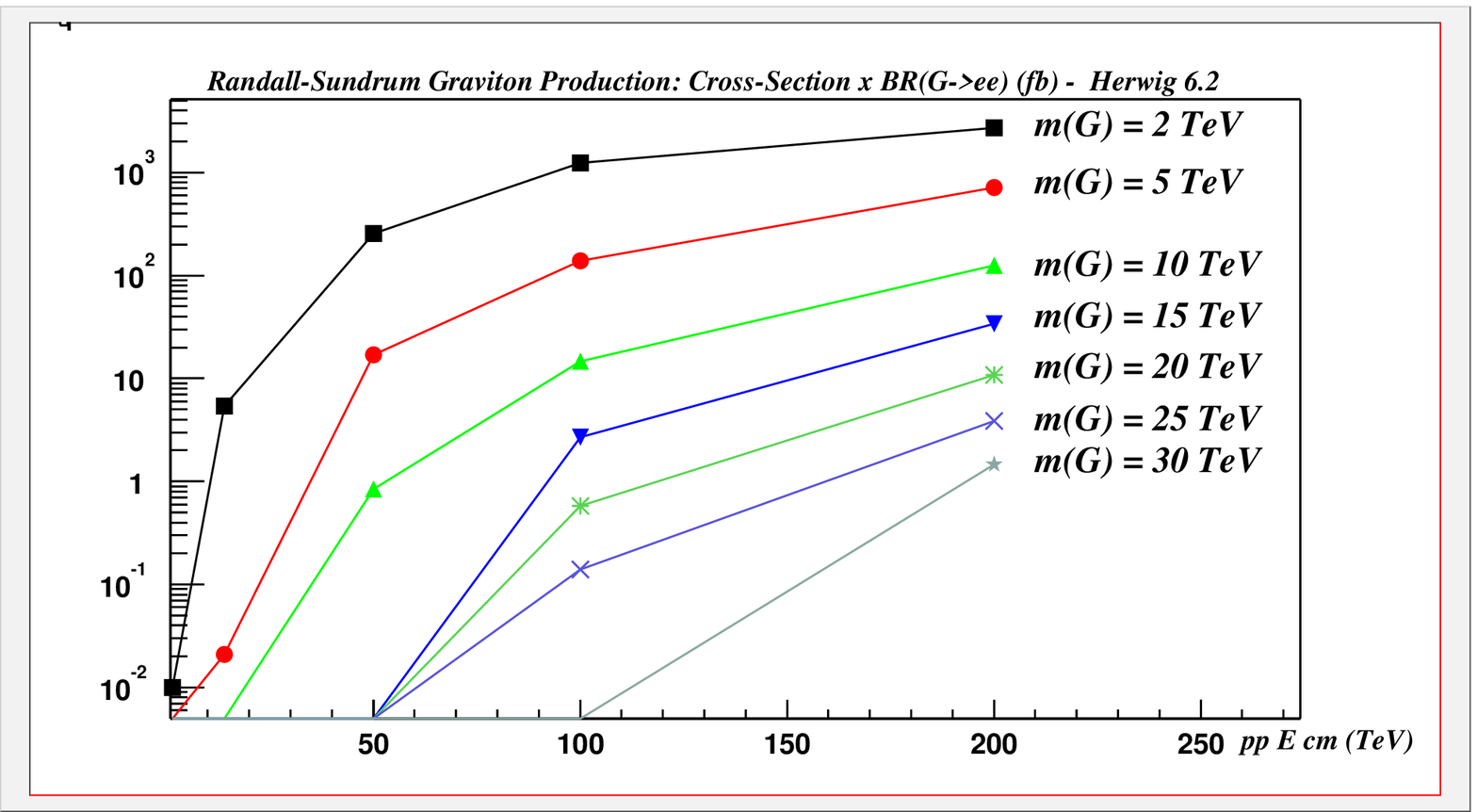}
    \vspace{0.45cm}
    \caption{$\sigma(pp\to G)\times BR(G\to e^+e^-)$ for various
graviton masses in models with warped extra dimensions as a function of the
center of mass energy for $\Lambda=10$~TeV.}
    \vspace{0.7cm}
    \label{fig:twentytwo}
  \end{minipage}
\end{figure}

For two or
more extra dimensions, the masses and couplings of KK excitations become
both level and compactification scheme dependent. This leads to a rather
complex KK spectrum in processes such as Drell-Yan production. It will
be necessary to observe a rather large part of the spectrum in order to
experimentally determine the number of extra dimensions and how they are
compactified. Some sample KK excitation spectra for a number of
different TeV-scale models with more than one extra dimension are shown
in Fig.~\ref{fig:twentyone}. The models are labeled by the manifold on
which they are compactified.
The spectra in these models are quite distinctive. Measuring the
location of the peaks, and their relative heights and widths can be used
to uniquely identify a given compactification scheme. 
Figure~\ref{fig:twentyone} suggests that one should be able to
differentiate the many possible models for compactification scales up to
$M_c\approx 20$~TeV through detailed cross section measurements. 

\subsubsection{Warped Extra Dimensions}

In models with warped extra dimensions, one expects to produce TeV-scale
graviton resonances in many channels, including the di-lepton
channel~\cite{rsphen}. In its simplest version, with one extra
dimension, two distinct branes, and with all SM fields on the TeV-brane,
this model has only two fundamental parameters which can be chosen to be
the mass of the first KK state, $m_1$, and the scale 
\begin{equation}
\Lambda=\bar M_{Pl}e^{-kr_c\pi},
\end{equation}
where $\bar M_{Pl}$ is the reduced effective 4-D Planck scale, $r_c$ is
the compactification radius of the extra dimension, and $k$ is a scale
of the order of the Planck scale. The masses of the graviton resonances
are given by
\begin{equation}
m_n=x_n(c)\Lambda
\end{equation}
where $c=k/\bar M_{Pl}$ and $x_n$ are the roots of the Bessel functions
of order~1, $J_1$. The mass of the first excitation is 
$m_1\approx 3.83~c\Lambda$. The 
couplings of the massive graviton resonances are given by
$1/\Lambda$. The decay angular
distribution in the resonance region can demonstrate that a spin-2
particle is being produced. Measurements of the relative branching ratios
to other clean decay modes such as $G\to\gamma\gamma$ can prove that one
is indeed producing gravitons. Table~\ref{tab:three} lists the branching
ratios for a few interesting decays for $m_1=2$~TeV. Detailed studies have 
shown~\cite{webber} that, at the LHC with an integrated luminosity of
100~fb$^{-1}$, it will be possible to discover graviton resonances with
masses up to $m_1=2.1$~TeV and to discriminate a spin-2 from a spin-1
resonance at 90\% CL for masses up to 1.7~TeV. 
The cross section for the production of the lightest
graviton resonance in $pp\to G\to e^+e^-$ as a function of $\sqrt{s}$ for
$\Lambda=10$~TeV and various graviton masses $m(G)=m_1$ is shown in 
Fig.~\ref{fig:twentytwo}. All cross sections were determined using {\tt
HERWIG6.2}~\cite{herwig}, which uses an essentially model independent
implementation of the resonance, only depending on the (universal)
coupling to the SM fields. At
the VLHC-II, with $\sqrt{s}=200$~TeV and an integrated luminosity of 
1~ab$^{-1}$, it should be possible to observe graviton
resonances with masses up to $m_1\sim 20$~TeV. 
\begin{table}
\caption{\label{tab:three} Branching ratios for a 2~TeV graviton resonance
for a few interesting decay modes.}
\begin{center}
\begin{tabular}{|c|c|}
\hline
Decay Mode & Branching Ratio ($M_G = 2\ TeV$) \\
\hline
$e^+e^-$ & 2.1 \% \\
$\bar tt$ & 6.1 \% \\
$\gamma\gamma$ & 4.2 \% \\
$W^+W^-$ & 9.0 \% \\
$ZZ$ & 4.5 \% \\
\hline
\end{tabular}
\end{center}
\end{table}

\subsubsection{Black Hole Production at the VLHC}

Should the LHC discover signals of extra dimensions such as KK
excitations of gravitons or of the SM gauge bosons, the VLHC might well be
able to access mass scales considerably larger than the fundamental (higher
dimensional) Planck scale, $M_P\sim$~TeV. A particularly
exciting consequence of this scenario is the production of black holes
(BHs)~\cite{bhprod,bhprod1}. Simple estimates~\cite{bhprod1} of their  
production cross section, treating the BHs as general relativistic
objects, suggest enormous event rates at the VLHC, perhaps as large as
$\sim kHz$ ($\sim 1Hz$) for $M_{BH}=10$~TeV ($M_{BH}=50$~TeV). 

Once produced, black holes decay primarily via Hawking radiation. The
decay of a BH is thermal; it obeys all local conservation laws but
otherwise does not discriminate between particle species (of the same
mass and spin). BHs thus decay with roughly equal probability to all of
the $\approx 60$ particles of the SM. The branching ratio of BHs into
leptons thus is about 10\%. Approximately 3\% of its decays result in SM
gauge bosons, 5\% in neutrinos, and 2\% in Higgs bosons in the final
state. The rest of $\sim 
80\%$ yields hadrons. The number of particles in the final state is
typically of ${\cal O}(10)$ and increases rapidly with $M_{BH}/
M_P$. The particles originating from BH decay each carry an energy of
several hundred GeV on the average. BH decays resulting in leptons or
photons in the final state thus result in clean signals with small SM
backgrounds. 

Since one expects only about 5\% missing transverse energy per event, it
should be possible to estimate $M_{BH}$ from the visible decay
products. The Hawking temperature, $T_H$, can be determined from the
energy spectrum of the final states. One can thus directly test the
hypothesis that the observed events originate from BH production and not from
other new physics. Furthermore, knowing $T_H$ as a function of $M_{BH}$
provides a tool to determine the number of spatial dimensions. $T_H$ and
$M_{BH}$ are related by
\begin{equation}
\log(T_H)=-{1\over n+1}\,\log(M_{BH})+{\rm const.}
\end{equation}

\subsection{Summary of Reach: A VLHC Pocket Guide}

The physics reach of the VLHC is summarized and compared with that of
the LHC and a luminosity or energy upgraded LHC in
Table~\ref{tab:threea}. 
\begin{table}
\caption{Search reach of the LHC, the SLHC and the VLHC for various new
physics scenarios. }
\vspace{2.mm}
\label{tab:threea}
\begin{tabular}{|c|c|c|c|c|c|}
\hline
physics & LHC & SLHC  & SLHC & VLHC-I  & VLHC-II \\
scenario & 100~fb$^{-1}$ & 14~TeV, 1~ab$^{-1}$ & 28~TeV, 100~fb$^{-1}$ &
40~TeV, 100~fb$^{-1}$ & 200~TeV, 100~fb$^{-1}$ \\
\hline
$\bar ttH$ coupling & 13\% & $\sim 10$\% & $\sim 7$\% & $5-10\%$ & $1-3\%$ \\
$M_{\bar g}$, $M_{\bar q}$ & 2~TeV & 2.5~TeV & $3-4$~TeV & $4-5.5$~TeV & $\sim
20$~TeV \\
mess. field $M_\Phi$ & -- & -- & -- & -- & $\sim 12$~TeV \\
strong $WW$ scat. & $1.7\,\sigma$ & $1.6\,\sigma$ & $4.5\,\sigma$ &
$7\,\sigma$ & $18\,\sigma$ \\
$M_{Z'}$ & $4-5$~TeV & $5-6$~TeV & $7-9$~TeV & $10-13$~TeV & $30-40$~TeV \\
comp. scale $\Lambda$ & 23~TeV & 35~TeV & 35~TeV & $\sim 50$~TeV & $\sim
100$~TeV \\
$M_{q^*}$ & 6.5~TeV & 7.5~TeV & 10~TeV & 13~TeV & $70-75$~TeV\\
extra dim., $\delta=2$, $M_D$ & 9~TeV & 12~TeV & 15~TeV & 24~TeV &
65~TeV\\
extra dim., $\delta=4$, $M_D$ & 6~TeV & 7~TeV & 10~TeV & 15~TeV & 45~TeV\\
\hline
\end{tabular}
\end{table}
The search reach of a SLHC with 1~ab$^{-1}$ is typically $20-30$\%
higher than that of the LHC. Instead, doubling the energy of the LHC
improves the reach by a factor~1.5 --~2. In a staged approach to a VLHC,
the first stage with $\sqrt{s}=40$~TeV would be able to find new
particles which are a factor 2~--~3 more massive than those which can be
accessed at the LHC. At the second stage, with $\sqrt{s}=200$~TeV, the
reach of the LHC can be improved by up to one order of magnitude. Note
that Table~\ref{tab:threea} is not exhaustive; only selected cases are
shown. 

%: section: detectors
\section{Detectors for Very High Energy Hadron Colliders}
\label{sec:three}

Given experiences 
at CERN and Fermilab, a broadly based assault on physics beyond the SM 
may well still be best probed with 
``general purpose'' ($GP$) detectors (although a need for specialized 
devices may indeed grow out of the physics results from the LHC). In the
following we discuss general requirements, energy deposition and
radiation levels, central tracking and outer muon tracking, and
calorimetry for such detectors at the VLHC, assuming luminosities in
the range ${\cal L}=10^{34} - 10^{35}\,{\rm cm^{-2}\,s^{-1}}$. A 
general conclusion (which deserves further thought)
was reached that  for  a
VLHC-I operating at $\sqrt{s}=40-50$~TeV and ${\cal L}=10^{34}\,{\rm
cm^{-2}\,s^{-1}}$, the technology of the current LHC detectors
ATLAS~\cite{atlas} and
CMS~\cite{cms} appears to be sufficient. A secondary conclusion was
that, for the machine 
parameters considered, there were no circumstances in which a $GP$-like 
detector could obviously not be built.

The objects to be identified and measured are 
jets, $\gamma$, $e$, $\mu$, $\tau$, charged tracks, and
$E\!\!\!\!/_T$~\cite{anderson}.
The $GP$ detectors will probably not be designed
with a capability of measuring or identifying individual hadrons. 
Magnetic analysis is mandatory for 
charge measurement, momentum analysis, muon identification 
and $b$-quark measurements.
%A likely combination is a 
%large inner solenoid combined with forward and outer toroids.
Generic tracking (in front of magnet and calorimetry) is required for
$e, \mu,\tau, b$ and may be provided by silicon (or eventually diamond)
microstrips, pixels or ``3D-pixels''. Tagging $b$-jets is 
important for Higgs boson, top and SUSY physics, which implies micro-vertexing
at the 100~$\mu$m level and the possibility of detecting $e$ and $\mu$
in jets. 

Electromagnetic (em) calorimetry is essential for both $e$- 
and $\gamma$-identification and measurement as well as measuring well
the electromagnetic  
component of jets; it is likely to have one or more ``shower maximum"
layers with very high granularity. Hadronic calorimetry is 
vital for jet spectroscopy ($t \rightarrow jjj, j\ell\nu$), QCD
measurements, as well as many searches for new physics, such as for 
compositeness, or black hole production. One needs to measure well the
direction (the core of the jets) and total energy, as independent as possible
of the em:hadronic ratio in the jet. Good hadron calorimetry, with a high 
dynamic range and good granularity, is also helpful in identifying and 
measuring (isolated) muons. Aggressive rapidity coverage of
the forward hadronic calorimeter is important for jet tagging which
plays a crucial role in all weak boson fusion processes (Higgs boson
production, weak boson scattering). As shown Sec.~\ref{sec:tagging1}, a
significant fraction of the tagging jets for $\sqrt{s}=200$~TeV is
produced with rapidities $|\eta(j)|>5$.

Measuring muons is critical. Many
massive fundamental objects ($W,Z$, SUSY, BH) decay to muons and one
needs to know
both their charge and their momentum. In addition, if muons are 
measured, one can determine $E\!\!\!\!/_T$. The resolution needed is
estimated to be $\delta p/p < 20$\% at 5~TeV. The best plan
is probably to \emph{measure} inside the central tracker well, and
\emph{verify} 
the muonic signature behind the calorimeter with relatively simple tracking.
Note that the sagitta of a 5~TeV particle over 2~m in a 4~T field is 
120~$\mu$m. This may be the driver on the central tracking resolution.

The necessity of measuring all the above objects can be illustrated by the
discovery of the top quark by CDF and D\O. Because the top cascades
($t \rightarrow bW, W \rightarrow e \nu,\mu\nu, \tau\nu, jj$) all the above
(except for $\gamma$ and $\tau$ detection) were essential. $\tau$
identification could be important 
for Higgs physics and testing lepton universality, e.g. in the case of a
massive $Z'$, and for supersymmetry.

Backgrounds and radiation loads are important for detector design, in
particular for luminosities of ${\cal L}=10^{35}\,{\rm cm^{-2}\,s^{-1}}$
or more. As total and inelastic cross sections increase only slowly
with energy
one can extrapolate with some confidence~\cite{superdet}. Most of the
produced hadrons 
have rather low transverse momentum $\langle p_T\rangle \approx 0.6$~GeV. The
radiation dose in the central region
is a function mostly of the luminosity, not the energy. This is not true in
the very forward region where particle momenta scale with beam energy.
We note that multiplicity distributions also widen only slowly with energy
but have long non-Gaussian tails. 
{\tt DPMJET/MARS} simulations have been carried out to estimate fluxes
of particles~\cite{dpm,mars}. At a VLHC-I (VLHC-II) the central tracker
will see, at $R = 10$~cm, $3(10) \times 10^7$ charged particles per
cm$^{-2}$ s$^{-1}$ and a tenth as many neutrons. This corresponds to
10(30)~MRad/year. 
At small polar angles (at an end-cap close to the beam pipe) one expects
100 -- 1000 times these fluxes! This is a major issue. Forward muon systems
\emph{must} have well-designed shielding, and then radiation damage
could be reduced to an acceptable level. Backgrounds from beam
losses are only a few percent 
of backgrounds from collisions, provided one has appropriate
shielding~\cite{strotenko}.

A number of physics processes, such as Higgs boson pair production (see 
Sec.~\ref{sec:2higgs}), or the production of messenger fields in gauge
mediated supersymmetry breaking models (see Sec.~\ref{sec:messenger})
would benefit from a luminosity of ${\cal L}=10^{35}\,{\rm cm^{-2}\,s^{-1}}$
or more. Current VLHC machine studies~\cite{m4talk} do not exclude this
possibility. For such high luminosities, one expects a very large
number of interactions per crossing, unless the bunch spacing is greatly
reduced compared to that of the LHC. A large number of interactions per
crossing leads to a large number of charged tracks and large
underlying transverse momenta which make isolation of particles difficult
and severely affect jet identification at all but the highest
energies. The number of underlying events, together with the number of
underlying charged tracks per rapidity unit, and the $E_T$ in a cone of
size $\Delta R=0.25$ for a number of cases are shown in
Table~\ref{tab:four}. 
%For comparison, results are also shown for Pb-Pb collisions at the LHC. 
%
\begin{table}
\caption{The average number of underlying events, ${\cal N}$, the
average number of underlying charged tracks per rapidity unit, $N_{tr}$,
and the total transverse energy, $E_T^{tot}$, in a cone of
size $\Delta R=0.25$ at the LHC and VLHC-II for several choices of the
instantaneous luminosity, ${\cal L}$, and the bunch spacing, $\Delta\tau$.\\ }
\label{tab:four}
\begin{tabular}{|c|c|c|c|c|c|}
\hline
$\sqrt{s}$ & ${\cal L}$ & $\Delta\tau$ & ${\cal N}$ & $N_{tr}$ & $E_T^{tot}$
\\
\hline
14~TeV & $10^{34}~{\rm cm^{-2}s^{-1}}$ & 25~ns & 20 & $\sim 160$ &
7.6~GeV \\
14~TeV & $10^{35}~{\rm cm^{-2}s^{-1}}$ & 12.5~ns & 100 -- 200 & $\sim
800 - 1600$ &
$38-76$~GeV \\
%14~TeV & Pb-Pb & 125~ns & (1) & $\sim 2500$ & ? \\
\hline
200~TeV & $10^{34}~{\rm cm^{-2}s^{-1}}$ & 18~ns & 24 & $\sim 240$ & 15~GeV \\
200~TeV & $10^{34}~{\rm cm^{-2}s^{-1}}$ & 6~ns & 8 & $\sim 80$ & 5~GeV \\
200~TeV & $10^{35}~{\rm cm^{-2}s^{-1}}$ & 18~ns & 240 & $\sim 2400$ &
150~GeV \\ 
200~TeV & $10^{35}~{\rm cm^{-2}s^{-1}}$ & 6~ns & 80 & $\sim 800$ & 50~GeV \\
%200~TeV & $10^{36}~{\rm cm^{-2}s^{-1}}$ & 18~ns & 2400 & $\sim 24000$ &
%1.5~TeV \\ 
%200~TeV & $10^{36}~{\rm cm^{-2}s^{-1}}$ & 6~ns & 800 & $\sim 8000$ & 0.5~TeV \\
\hline
\end{tabular}
\end{table}

\subsection{Tracking}

Tracking using gaseous drift cells larger than about 1~cm is 
too slow for the central tracker, but would be acceptable for the muon 
system. All silicon central tracking is an option, with strips or ministrips
(e.g. 50~$\mu{\rm m}\times5$~mm) or pixels of the pad variety or
columns. We consider
up to $10^9$ elements (40~layers about 4~m long out to 2~m radius). In
the column or 3D geometry electric field lines end on cylinders with axes
normal to the detector
plane, so the drift is transverse to the particle direction. One can then have
short collection distances and times (most of the signal is induced when the
charge is close to the electrode, where the electrode solid angle is large)
so the 3D signals are concentrated in time. Keys to the technology are being 
able to etch deep, near vertical holes and coat them with
polysilicon. The first
3D detectors have been successfully fabricated. There is a wide plateau 
and calculations indicate a pulse
duration of about 1~ns. R\&D on these column detectors includes developing
techniques for fabricating large areas, minimizing the amount of material 
for multilayer trackers, and studies of radiation hardness.

\subsection{Calorimetric Techniques}

Both electromagnetic and hadronic calorimeters are essential for measuring
electrons, photons and jets. One requires radiation hardness, high granularity,
and speed. The different technology choices with their advantages and
disadvantages should be clearly spelled out following on experience at 
Run~IIb and LHC.
%are summarized in Table~\ref{tab:five}.\marginpar{\tiny what does ``='' mean exactly?}
%
%\begin{table}
%\caption{Technology choices for calorimeters and their advantages ($+$)
%and disadvantages ($-$). }
%\label{tab:five}
%\vspace{1.mm}
%\begin{tabular}{|c|c|c|c|c|}
%\hline
%technology & Radiation hardness & Speed & Granularity & Cost \\
%\hline
%LAr & $+$ & = & = & = \\
%Scintillator & $+$ & = & = & ? \\
%Crystals & $+$ & = & $+$ & ?\\
%Quartz fibers & $+$ & ? & ? & ? \\
%Silicon pads & $-$ & ? & ? & ? \\
%Diamond pads & $+$ & ? & ? & $-$ \\
%High Pr. gas tubes & $+$ & ? & ? & $+$\\
%\hline
%\end{tabular}
%\end{table}
%
Scintillator read out with 
embedded wavelength-shifting fibers to a light detector (PMT, APD, HPD) is a
well-established technique which may be considered especially for the central 
region. Lead tungstate PbWO$_4$ or other crystals make a good electromagnetic 
calorimeter but compromise
the jet resolution. Quartz fibers embedded in (e.g.) Cu gives a low (Cerenkov)
signal but are very radiation hard. Silicon pads make a compact 
calorimeter but radiation hardness is an issue. (CVD) Diamond pads should be 
radiation hard enough but at present the cost would be prohibitive. This is
a clear R\&D issue: to bring down the cost of diamond films so that they can
be realistically considered for forward (if not central) calorimetry. High 
pressure gas calorimetry is cheap,
radiation hard and promising, especially for the forward direction.
Liquid Argon is intrinsically slow (compared
to 18~ns) but
with signal shaping and leading-edge recognition has been shown to be 
robust in a relatively high rate environment.
A particularly interesting new idea is to use unsegmented dual read-out 
calorimetry. In this concept, 
longitudinal quartz fibers, which see mostly the electromagnetic shower,
and scintillating fibers, which see mostly the hadronic shower, 
are embedded in a metal absorber matrix. 
This can achieve the essential advantages
of compensating calorimetry, eliminating the effect of fluctuations in 
$f_{em}$.
The sampling fraction could be as large as needed for optimizing the em 
response,
and the hadronic resolution would not suffer. It is an important R\&D 
project to construct and test such a dual readout calorimeter.

\subsection{Muons}

Muons are important for $W,Z,t$ and $b$-physics as well as knowing the 
missing $E_T$, $E\!\!\!\!/_T$. Because of multiple scattering and energy
loss fluctuations in the calorimeter, it is best to measure the muon
momentum before
the calorimeter, and use the track behind the calorimeter to identify the muon.
There are large fluctuations ($20-30$~GeV) in $p_{out}/p_{in}$ for a 1~TeV
muon~\cite{pdg}. A goal is to measure the sign of $\approx 10$~TeV
muons with 
a combination of field integral (8~Tm), number of points ($40-50$), and 
precision
per point (50~$\mu$m). The tracking outside the calorimeter, which is 
very large in area, can then be relaxed. When muons are isolated
one can and should measure any showering losses along their tracks, as these
are significant.

\subsection{Detectors for a VLHC: Findings and Conclusions}

In order to make optimal use of the enormous center of mass energy of a
VLHC, hadron calorimeter coverage out to $|\eta|=6-7$ and luminosities in
the range ${\cal L}=10^{34}-10^{35}\,{\rm cm^{-2}\,s^{-1}}$ are
necessary. At the upper end of this range, numerous technological
challenges are present for almost all detector components. From the 
very preliminary
survey~\cite{talks} described here, one can begin to form tentative
conclusions which 
should lead to specific R\&D efforts. These tentative conclusions are:
\begin{enumerate}
\item  Early indications suggest that with existing and anticipated 
technologies, tracking may be manageable at the necessary level. However,
a significant amount of R\&D
directed towards radiation hard, precise and fast approaches (e.g. silicon
pads, 3D pixels) is necessary to conclusively answer this question. 
\item It may be possible to identify, sign, and momentum analyze muons at the
necessary level for all luminosities. The main issue is the
momentum resolution for muons with momenta larger than 1~TeV. This needs to
be looked at in detail in the context of punch-through and backgrounds.
\item Central calorimetry will likely survive the anticipated radiation 
doses. However, the large number of interactions per crossing will make
jet identification and the measurement of missing transverse momentum
difficult, even in the multi hundred GeV region. The alternative of
extremely short bunch spacing represents major challenges for triggering
and data acquisition.
\item In the forward region, which is important for jet tagging in WBF
events, radiation hardness is a major issue. Here, a large scale R\&D effort
directed towards quartz fibers, scintillating fibers and silicon/diamond
pads is needed in order to see to what rapidities the hadron calorimeter
can be extended. 
\item A general conclusion reached was that if LHC detectors function as
expected at $10^{34}\,{\rm cm^{-2}\,s^{-1}}$
then scaled detectors with ATLAS/CMS technologies would also work at
VLHC-I ($\sqrt{s}=40-50$~TeV) at the same luminosity.
\end{enumerate}

In an important sense the major R\&D for VLHC detectors is the
construction and successful use of
LHC detectors ATLAS and CMS. If the LHC detectors were scaled up a 
factor~1.4 linearly (2.7 in volume) that
would be a fair starting point for a VLHC-II  at $2\cdot 
10^{34}\,{\rm cm^{-2}\,s^{-1}}$. 
But for all detectors we need more radiation hard technology, we need more
precision/granularity, and we generally need them to be faster. These are the
areas of R\&D needed, together with finding ways of making the detectors
cheaper.
It is not unreasonable to suppose that we have some 10~years (until $\approx
2010$) for the R\&D and then one will have 10~years to build the detectors,
for operation $\approx 2020$. 

\section{Conclusions and Opinions}
\label{sec:four} 

In this report we have presented preliminary results of a
survey of the physics capabilities of a hadron collider operating in the
100~TeV region. We have also studied the detector requirements so
that the physics goals can be achieved. 

If the Higgs boson is light ($M_H\leq 200$~GeV), as indicated by present
precision electroweak data, a LC will be an ideal tool to precisely
study the properties of such a particle. No such ``gold plated'' physics
case exists at this point for a VLHC. However, although the LHC has
excellent prospects of making fundamental discoveries beyond the SM, it
is unlikely to provide complete answers to all ensuing questions. 
It is easy to construct
scenarios where a convincing reason for building a VLHC emerges. In the
following we list some which directly follow from the topics discussed
in Sec.~\ref{sec:two}.
\begin{itemize}
\item The Tevatron and/or LHC/SLHC finds a Higgs boson and nothing else. A LC
will then measure most of the Higgs boson properties if $M_H$ is
sufficiently small. The VLHC
would extend those measurements and be the only device capable of 
searching for the necessary next scale.
\item The Tevatron and/or LHC/SLHC finds a light Higgs boson consistent with
a supersymmetric interpretation and nothing else. This could happen for
certain points in parameter space of MSUGRA models. In this case the
first stage of the VLHC would find supersymmetry.
\item The LHC/SLHC finds some supersymmetric particles, but no squarks
associated with the first two generations. This situation could easily
happen in inverted mass hierarchy models. The VLHC would then find the
missing squarks and give important hints which model is realized in
nature. 
\item The LHC/SLHC finds supersymmetry and determines that the messenger
scale is of ${\cal O}(10-100$~TeV). In this case, the messenger fields
which communicate supersymmetry breaking would be directly produced at the
VLHC. 
\item The LHC/SLHC does not find anything except for a weak indication of an
enhanced rate in vector boson scattering at the highest accessible
energies. This could happen in certain models where the electroweak
symmetry is broken dynamically, such as the K-matrix unitarization
scheme. VLHC-I would see a clear signal.
\item The LHC/SLHC finds evidence for contact interactions with a scale
$\Lambda<60$~TeV. The VLHC would
then be able to directly probe the scale of new physics, eg. in
composite models of quarks and leptons, excited quarks would be
produced. For such measurements, VLHC-I, II are the obvious devices to 
exploit that finding.
\item The LHC/SLHC finds KK excitations of gravitons and/or the SM gauge
bosons. In this case the VLHC could perform detailed measurements of the
excitation spectrum. This would make it possible to determine how the
additional spatial dimensions associated with the KK excitations are
warped or compactified. 
\item If the  LHC/SLHC finds evidence for extra dimensions, the VLHC 
may well operate 
sufficiently far above the (higher dimensional) Planck scale such that
black holes are produced. BH production would make it possible to actually
measure the number of extra dimensions.
\end{itemize}

In all cases information obtained at the LHC/SLHC gives important clues
on what physics lies beyond the SM. However, it is not necessary to wait
for results from the LHC in order to start planning for a VLHC now. At
some point one, inevitably, will want to explore the multi
10~TeV region. A hadron collider is the only machine we in principal
know how to build which can directly discover new physics in this
region. With the long lead time for a large project like a
hadron collider it is essential to start this process now. Furthermore,
this must be done as part of a coordinated and coherent international
plan which is part of a comprehensive and global High Energy Physics
Program. 

At this workshop we have begun to investigate the physics potential of a
Very Large Hadron Collider with a center of mass energy up to
200~TeV. We have also identified several important areas of detector
R\&D for such a machine. In the three weeks available we have barely
scratched the surface of many topics. More serious and detailed studies
in the next several years are essential to fully unravel the potential of a
VLHC. 

%\begin{acknowledgments}
%\end{acknowledgments}

%: \begin{thebibliography}{3}

\end{document}